\documentclass[12pt,a4paper]{article}
\usepackage{jcappub}
\usepackage{amsmath}
\usepackage{latexsym}
\usepackage{amsfonts}
\usepackage{mathrsfs}
\usepackage{epstopdf}

\title{The reconstruction of tachyon inflationary potentials}

\author{Qin Fei,}
\author[1]{Yungui Gong,\note{Corresponding author}}
\author{Jiong Lin,}
\author{and Zhu Yi}
\affiliation{School of Physics, Huazhong University of Science and Technology,1037 LuoYu Rd,
Wuhan, Hubei 430074, China}

\emailAdd{feiqin@hust.edu.cn}
\emailAdd{yggong@mail.hust.edu.cn}
\emailAdd{707751841@qq.com}
\emailAdd{yizhu92@hust.edu.cn}

\abstract{
We derive a lower bound on the field excursion for the tachyon inflation, which is determined by the amplitude of the scalar perturbation
and the number of $e$-folds before the end of inflation. Using the relation between the observables like $n_s$ and $r$ with the slow-roll parameters,
we reconstruct three classes of tachyon potentials. The model parameters are determined from the observations before the potentials are reconstructed,
and the observations prefer the concave potential. We also discuss the constraints from the reheating phase preceding the radiation domination for
the three classes of models by assuming
the equation of state parameter $w_{re}$ during reheating is a constant. Depending on the model parameters and the value of $w_{re}$, the constraints on $N_{re}$ and $T_{re}$ are different.
As $n_s$ increases, the allowed reheating epoch becomes longer for $w_{re}=-1/3$, 0 and $1/6$
while the allowed reheating epoch becomes shorter for $w_{re}=2/3$.
}

\arxivnumber{1705.02545}

\begin{document}

\maketitle

\section{Introduction}

Inflation not only provides the solution to the monopole, horizon and flatness problems, but also provides the seeds for the large scale structure of the Universe.
The inflationary phase is usually driven by the potential or vacuum energy of a scalar field called the inflaton with a flat potential.
Motivated by string theory, the tachyon condensate with the effective
Dirac-Born-Infeld action is an interesting scalar field
and the cosmological consequences of the rolling tachyon were widely studied \cite{Sen:2002nu,Sen:2002in,Gibbons:2002md,Padmanabhan:2002cp,Frolov:2002rr}.
Tachyon inflation also provides the almost scale invariant power spectrum \cite{Garriga:1999vw,Hwang:2002fp,Steer:2003yu}.
To compare inflationary models with the observations, we need to calculate the observables $n_s$ and $r$ for a pivotal scale $k_*$, and the results
are usually expressed in terms of the number of $e$-folds $N_*$ before the end of inflation at the horizon exit of the pivotal scale.
For example, the chaotic inflation with the power-law potential $\phi^p$ gives $n_s=1-(p+2)/(2N_*)$ and $r=4p/N_*$ \cite{linde83},
the Starobinsky model gives $n_s=1-2/N_*$ and $r=12/N^2_*$ \cite{starobinskyfr} which is consistent with
the Planck 2015 results $n_s=0.9645\pm 0.0049$ and $r_{0.002}<0.10$ \cite{Ade:2015lrj}.
Therefore, we can parameterize the observables or the slow-roll parameters with $N$ for inflationary models \cite{Huang:2007qz, Gobbetti:2015cya}.
Furthermore, by parameterizing the slow-roll parameters or the observable $n_s$ with $N$, we can constrain the model parameters easily and reconstruct the inflationary potentials
\cite{Mukhanov:2013tua,Roest:2013fha,Garcia-Bellido:2014gna,Garcia-Bellido:2014wfa,Garcia-Bellido:2014eva,Creminelli:2014nqa,Boubekeur:2014xva,
Barranco:2014ira,Galante:2014ifa,Gobbetti:2015cya,Chiba:2015zpa,Cicciarella:2016dnv,Lin:2015fqa,Nojiri:2010wj,
Odintsov:2016vzz,Yi:2016jqr,Odintsov:2017qpp,Nojiri:2017qvx,Choudhury:2017cos,Gao:2017uja,Jinno:2017jxc,Gao:2017owg}.
For the ultra slow-roll inflation \cite{Tsamis:2003px,Kinney:2005vj}, the slow-roll parameter $\eta=\ddot\phi/(H\dot\phi)$ is a constant
and the reconstruction was discussed in \cite{Martin:2012pe,Motohashi:2014ppa,Motohashi:2017aob,Motohashi:2017vdc}.
The reconstruction method was also applied to tachyon inflation by parameterizing the slow-roll parameter $\epsilon$ or equivalently $r$ with $N$ \cite{Barbosa-Cendejas:2015rba}.

In addition to the constraints on $n_s$ and $r$, it was proposed that the reheating phase preceding
the radiation domination may provide further constraints on inflationary models \cite{Dai:2014jja}.
Assuming that the effective equation of state parameter $w_{re}$ is a constant, we can relate the total
number of $e$-folds during reheating with $N_*$ and the energy scale at the end of inflation \cite{Dai:2014jja,Cook:2015vqa,Ueno:2016dim,Kabir:2016kdh,DiMarco:2017sqo,Dimopoulos:2017zvq}.
In this paper, we use the reconstruction method by assuming either constant or simple inverse power-law parametrization to reconstruct tachyon potentials
and discuss additional constraints from reheating.

The paper is organized as follows. In section II, we review the tachyon inflation and the reconstruction method.
The lower bound on the field excursion are derived, and the relation between the reconstruction method and the generalized $\beta$-function method
is also discussed. In section III, we reconstruct the classes of potentials for the constant $\eta_V$, the simple parametrization $n_s=1-p/(N+A)$
and the inverse power-law parametrization $r=16\gamma/(N+\alpha)^\beta$. By assuming that the equation of state parameter $w_{re}$ during reheating is a constant,
we discuss the constraints on reheating for the three models in section IV, the paper is concluded in section V.

\section{Tachyon inflation}

For more general scalar fields, the kinetic term may not take the standard canonical form. In particular,
the tachyon condensate in the string theory can be described by an effective scalar field with nonlinear kinetic term
which drives inflation even without the help of the potential.
The effective action for the rolling tachyon is
\begin{equation}
\label{tachyon1}
S_T=-\int d^4x \sqrt{-g}\,V(T)\sqrt{1+g^{\mu\nu}\partial_\mu T\partial_\nu T}.
\end{equation}
Applying the Arnowitt-Deser-Misner (ADM) metric \cite{adm},
\begin{equation}
\label{eq2}
d{s^2} =  - {\mathscr{N}^2}d{t^2} + {h_{ij}}\left( {d{x^i} + {N^i}dt} \right)\left( {d{x^j} + {N^j}dt} \right),
\end{equation}
the gravitational and tachyon action becomes
\begin{equation}
\label{tachyon2}
S=\int d^4x \mathscr{N}\sqrt h\left\{\frac{1}{2}\left[^{(3)}R+\frac{1}{\mathscr{N}^2}(E^{ij}E_{ij}-E^2)\right]-V\left(1-\frac{\dot T^2}{\mathscr{N}^2}\right)^{\frac{1}{2}}\right\},
\end{equation}
where $\mathscr{N}$ and $N^i$ are the lapse and the shift functions, respectively,
all the spatial indices are raised
and lowered by the metric $h_{ij}$ for the three dimensional space, $\dot T=dT/dt$,
\begin{equation}
\label{eq4}
{E_{ij}}=\frac{1}{2}\left( {{{\dot h}_{ij}} - {\nabla _i}{N_j} - {\nabla _j}{N_i}} \right),
\end{equation}
$E=h^{ij}E_{ij}$, the extrinsic curvature $K_{ij}=E_{ij}/\mathscr{N}$, and
the covariant derivative is with respect to the three dimensional spatial metric $h_{ij}$. Note that
we take $M_{pl}=1/\sqrt{8\pi G}=1$.
Since the lapse and shift functions $\mathscr{N}$ and $N_i$
contain no time derivative, the variations with respect to them give the
corresponding Hamiltonian and momentum constraints,
\begin{equation}
\label{pconstraint1}
\nabla_i\left[\frac{1}{\mathscr{N}}(E^i_j-h_{ij} E)\right]=0,
\end{equation}
\begin{equation}
\label{econstraint1}
^{(3)}R-\frac{1}{\mathscr{N}^2}(E^{ij}E_{ij}-E^2)-2V\left(1-\frac{\dot T^2}{\mathscr{N}^2}\right)^{-\frac{1}{2}}=0.
\end{equation}

For the homogeneous and isotropic background,
$\mathscr{N}=1$, $N_i=0$ and $h_{ij}=a^2\delta_{ij}$,
the Hamiltonian constraint (\ref{econstraint1}) becomes the Friedmann equation
\begin{equation}
\label{frweq1}
H^2=\frac{1}{3}\frac{V}{\sqrt{1-\dot T^2}},
\end{equation}
and the momentum constraint satisfies automatically.
The energy density and the equation of state for the tachyon field are
\begin{gather}
\label{tacheq1}
\rho=\frac{V}{\sqrt{1-\dot{T}^2}},\\
w=\frac{p}{\rho}=\dot{T}^2-1.
\end{gather}
The equation of motion for the tachyon field is
\begin{equation}
\label{frweq2}
\frac{\ddot T}{1-\dot T^2}+3H\dot T+\frac{V_{,T}}{V}=0,
\end{equation}
where $V_{,T}=dV/dT$.
Combining eqs. \eqref{frweq1} and \eqref{frweq2}, we get
\begin{equation}
\label{acceq1}
\dot H=-\frac{3}{2}H^2\dot T^2.
\end{equation}
Before we review the slow-roll inflation and perturbations, we discuss the inflationary attractor \cite{Liddle:1994dx} first.
Combining eqs. \eqref{frweq1} and \eqref{acceq1}, we get the Hamilton-Jacobi equation
\begin{equation}
\label{H-J}
V^2=9H^4-4(dH/dT)^2.
\end{equation}
Suppose $H_0(T)$ is a solution to eq. \eqref{H-J}, either inflationary or noninflationary, then we consider a perturbation $\delta H(T)$ around $H_0(T)$,
i.e., another trajectory $H_0(T)+\delta H(T)$ which satisfies eq. \eqref{H-J}.
To the linear perturbation, we get
\begin{equation}
\label{p:H-J}
9H_0^3\delta H=2(dH_0/dT)(d\delta H/dT).
\end{equation}
In terms of the number of e-folds $N$ before the end of inflation, we get the solution
\begin{equation}
\label{deltahsol}
\delta H(T)=\delta H(T_i)\exp\left[-3\left(N_i-N\right)\right].
\end{equation}
If $H_0(T)$ is an inflationary solution, then all linear perturbations approach it exponentially as the tachyon rolls down,
so inflationary attractor exists if the potential is able to support inflation.

\subsection{slow-roll inflation}

From eq. \eqref{acceq1}, we get
\begin{equation}
\label{acceq2}
\frac{\ddot a}{a}=\dot H+H^2=H^2\left(1-\frac{3}{2}\dot T^2\right).
\end{equation}
The condition for inflation $\ddot a>0$ requires $\dot T^2<2/3$. If the tachyon field satisfies
the slow-roll conditions
\begin{gather}
\label{slreq1}
\dot T^2\ll 1,\\
\label{slreq2}
\quad \ddot T\ll 3H\dot T,
\end{gather}
then the background equations during inflation are
\begin{gather}
\label{frweq3}
H^2\approx \frac{V}{3}, \\
\label{frweq4}
3H \dot T\approx -V_{,T}/V.
\end{gather}
By using the number of $e$-folds $N(t)=\ln (a_e/a)$ before the end of inflation,
\begin{equation}
\label{neeq1}
N(t)=\int_t^{t_e} H(t)dt,
\end{equation}
we introduce the horizon-flow slow-roll parameters \cite{Schwarz:2001vv}
\begin{gather}
\label{slreq3}
\epsilon_0=\frac{H_*}{H},\\
\label{slreq4}
\epsilon_{i+1}=-\frac{d \ln|\epsilon_i|}{dN},
\end{gather}
where the subscript $e$ denotes the end of inflation, the subscript $*$ denotes the horizon crossing
and we choose $H_*$ as the Hubble parameter at the horizon crossing
for a particular scale, for example, $k_*=0.002$ Mpc$^{-1}$. For the tachyon field,
the first two slow-roll parameters are \cite{Steer:2003yu}
\begin{gather}
\label{slreq5}
\epsilon=\epsilon_1=-\frac{\dot H}{H^2}=\frac{3}{2}\dot T^2\approx \frac{1}{2}\frac{V^2_{,T}}{V^3},\\
\label{slreq6}
\eta=\epsilon_2=2\frac{\ddot T}{H\dot T}\approx -2\frac{V_{,TT}}{V^2}+3\frac{V^2_{,T}}{V^3}.
\end{gather}
From eqs. \eqref{neeq1} and \eqref{slreq5}, we get
\begin{equation}
\label{neeq2}
N(T)=\pm \sqrt{\frac{3}{2}}\, \int_T^{T_e}\frac{H}{\sqrt{\epsilon}}dT\approx \int^T_{T_e} \frac{V^2}{V_{,T}} dT,
\end{equation}
where the $\pm$ sign is the same as the sign of $\dot T$.

\subsection{Perturbations}

For convenience, we choose the flat gauge,
\begin{equation}
\label{eq13}
\begin{split}
\delta T(x,t) = 0,\ \mathscr{N}=1+N_1,  \ N_i=\partial_i\psi+ N_i^T,\\
h_{ij} = a^2 \left((1 + 2\zeta  + 2\zeta ^2)\delta_{ij} + \gamma _{ij} + \frac{1}{2}\gamma _{il}\gamma _{lj} \right),
\end{split}
\end{equation}
where ${\partial^i}N_i^T = 0$, $\zeta$ and $\gamma_{ij}$ denote the scalar and tensor fluctuations respectively,
the tensor perturbation satisfies
${\partial_i}{\gamma^{ij}} = 0$ and ${h^{ij}}{\gamma _{ij}} = 0$.
Note that $N_1$, $\psi$, $N_i^T$, $\zeta$ and $\gamma_{ij}$ are first order quantities.
Substituting eq. \eqref{eq13} into the momentum constraint \eqref{pconstraint1} and the Hamiltonian constraint \eqref{econstraint1},
to the first order, we get the solution $N_i^T=0$ and
\begin{equation}
\label{eq16}
\begin{split}
N_1&= \frac{\dot \zeta }{H},\\
\psi&=-\frac{\zeta}{a^2 H}+\chi,\\
\nabla^2 \chi&=\frac{3}{2}\frac{\dot T^2}{1-\dot T^2}\dot\zeta.
\end{split}
\end{equation}
Combining the solution \eqref{eq16} with the background equations \eqref{frweq1}, \eqref{frweq2} and \eqref{acceq1},
to the second order of perturbation, the action \eqref{tachyon2} for the scalar perturbation becomes
\begin{equation}
\label{tachyon3}
S=-\frac{3}{2}\int d^4x \left[a\dot T^2(\partial_i\zeta)^2-a^3\frac{\dot T^2}{1-\dot T^2}\dot\zeta^2\right].
\end{equation}
Using the canonically normalized field $v = z \zeta$, where
\begin{equation}
\label{normvars}
z = \frac{\sqrt 3a\dot T}{\sqrt{1-\dot T^2}},
\end{equation}
the action (\ref{tachyon3}) becomes
\begin{equation}
\label{tachyon4}
S = \int d^3x d\tau \frac{1}{2} \left[ v{'^2} - c_s^2 (\partial_i v)^2 + \frac{z''}{z} v^2 \right],
\end{equation}
where the prime denotes the derivative with respect to the conformal time $\tau=\int dt/a$,
and the effective sound speed is $c_s^2 = 1-\dot T^2$ \cite{Garriga:1999vw}. In terms of the slow-roll parameters,
we get \cite{Steer:2003yu}
\begin{equation}
\label{slreq7}
\frac{z''}{z}\approx a^2H^2(2-\epsilon+\frac{3}{2}\eta).
\end{equation}
To discuss the quantum fluctuations, we define the operator
\begin{equation}
\label{veq21}
\hat v(\tau ,\vec{x}) = \int \frac{d^3k}{( 2\pi)^3}\left[v_k(\tau)\hat a_k e^{i\vec{k} \cdot \vec{x}}+v_k^*(\tau)\hat a_k^\dag e^{-i\vec{k} \cdot \vec{x}}\right],
\end{equation}
where the creation and annihilation operators satisfy the standard commutation relations
\begin{equation}
\label{veq22}
\begin{split}
\left[\hat a_k,\ \hat a_{k'}^\dag\right]=(2\pi)^3\delta^3(\vec{k}-\vec{k'}),\\
\left[\hat a_k,\ \hat a_{k'}\right]=\left[\hat a_k^\dag,\ \hat a_{k'}^\dag\right]=0,
\end{split}
\end{equation}
and the mode functions obey the normalization condition
\begin{equation}
\label{eq23}
v_k' v_k^* - v_k {v_k^*}' =  - i.
\end{equation}
We choose the Bunch-Davis vacuum defined by $\hat a_k|0\rangle=0$.
Varying the action \eqref{tachyon4} and using eq. \eqref{slreq7},
we obtain the Mukhanov-Sasaki equation for the mode function $v_k(\tau)$ \cite{Steer:2003yu},
\begin{equation}
\label{eq21}
v_k'' + \left( c_s^2 k^2 - \frac{\nu^2-1/4}{\tau^2} \right)v_k = 0,
\end{equation}
where
\begin{equation}
\label{slreq8}
\nu=\frac{3}{2}+\epsilon+\frac{1}{2}\eta.
\end{equation}
Solving eq. \eqref{eq21} with the condition \eqref{eq23}, we find that outside the horizon,
the scalar perturbation is almost a constant,
\begin{equation}
\label{vksol8}
|\zeta_k|=\frac{|v_k|}{z}=2^{\nu-\frac{5}{2}}\frac{\Gamma(\nu)}{\Gamma(3/2)}
\frac{H(1-\epsilon)^{\nu-1/2}}{c_s^{1/2} k^{3/2} \epsilon^{1/2}}\left(\frac{c_s k}{aH}\right)^{\frac{3}{2}-\nu}.
\end{equation}
Therefore, the power spectrum of the scalar perturbation is \cite{Steer:2003yu}
\begin{equation}
\label{pkeq1}
P_{\zeta}=\frac{k^3}{2\pi^2}|\zeta_k|^2=\left[1-\left(\frac{5}{3}+2C\right)\epsilon-C\eta\right]\left.\frac{H^2}{8\pi^2\epsilon}\right|_{c_sk=aH},
\end{equation}
where $C=\gamma+\ln2-2\approx-0.72$. The amplitude of the scalar perturbation is
\begin{equation}
\label{aseq1}
A_s=\left.\frac{H^2}{8\pi^2\epsilon}\right|_{c_sk=aH}.
\end{equation}
The scalar spectral tilt is \cite{Garriga:1999vw,Steer:2003yu}
\begin{equation}
\label{nseq1}
n_s-1=\left. \frac{d\ln P_\zeta }{d\ln k} \right|_{c_sk = aH}=-2\epsilon-\eta.
\end{equation}

For the tensor perturbation, to the second order, the action \eqref{tachyon2} becomes
\begin{equation}
\label{tachyon7}
S=\frac{1}{8}\int d^4x\left[a^3(\dot\gamma_{ij})^2-a(\gamma_{ij,k})^2\right],
\end{equation}
so the tensor spectrum is \cite{Steer:2003yu}
\begin{equation}
\label{pteq1}
P_T=[1-2(1+C)\epsilon]\left.\frac{2H^2}{\pi^2}\right|_{k=aH}.
\end{equation}
The tensor spectral tilt is \cite{Steer:2003yu}
\begin{equation}
\label{nteq1}
n_T=-2\epsilon[1+\epsilon+(1+C)\eta].
\end{equation}
The tensor to scalar ratio is \cite{Garriga:1999vw,Steer:2003yu}
\begin{equation}
\label{req1}
r=16\epsilon\left[1-\frac{1}{3}\epsilon+C\eta\right].
\end{equation}

\subsection{Field excursion}
If $\epsilon$ is a monotonic function and $H$ decreases during inflation, then from eqs. \eqref{neeq2} and \eqref{aseq1},
we get
\begin{equation}
\label{gfbld1}
N_*\le \sqrt{\frac{3}{2}}\, \frac{H_*}{\sqrt{\epsilon(T_*)}}|T_e-T_*|=\sqrt{12\pi^2 A_s}\, M_{pl}\Delta T.
\end{equation}
In the last equality, we write out $M_{pl}$ explicitly. Therefore, similar to the Lyth bound \cite{Lyth:1996im,Gao:2014pca},
there is a lower bound on the field excursion for the tachyon,
\begin{equation}
\label{gfbld2}
M_{pl}\Delta T\ge \frac{N_*}{\sqrt{12\pi^2 A_s}}\approx 1.18\times 10^5\left(\frac{N_*}{60}\right),
\end{equation}
where we use the observational value $\ln(10^{10} A_s)=3.094$ \cite{Ade:2015lrj}.

\subsection{Summary of the relations}
\label{reconsrel}

From eqs. \eqref{neeq1} and \eqref{slreq5}, we get
\begin{equation}
\label{slreq9}
\epsilon=\frac{3}{2}\dot T^2\approx\frac{V_{,N}}{2V}.
\end{equation}
From eq. \eqref{slreq6}, we get
\begin{equation}
\label{slreq10}
\eta=\frac{2\ddot T}{H\dot T}=-\frac{d\ln\epsilon}{dN}.
\end{equation}
Substituting eqs. \eqref{slreq9} and \eqref{slreq10} into eq. \eqref{nseq1}, we get
\begin{equation}
\label{nseq2}
n_s-1=-2\epsilon+\frac{d\ln\epsilon}{dN}=\left(\ln\frac{V_{,N}}{V^2}\right)_{,N}.
\end{equation}
From these relations, we see that once we parameterize one of the observable $n_s$ and $r$
or the slow-roll parameters $\epsilon$ and $\eta$ by $N$, we can derive all other parameters
and the potential $V(N)$.  Additionally, if we use the following relation
\begin{equation}
\label{tneq1}
dT\approx \pm \frac{\sqrt{V_{,N}}}{V} dN,
\end{equation}
to derive the function $T(N)$, then we can reconstruct the potential $V(T)$, where the $\pm$ sign is the same as the sign of $dV/dT$.
Alternatively, we can derive $V(T)$ with the following relation,
\begin{equation}\label{tneq2}
\frac{dV}{dT}=\frac{dV}{dN}\frac{dN}{dT}=\pm V^{3/2}(2\epsilon)^{1/2}.
\end{equation}
Therefore, by specifying one of the functions $\epsilon(N)$, $\eta(N)$, $n_s(N)$, $T(N)$ and $V(N)$,
we can derive the observable $n_s(N)$ and $r(N)$ and reconstruct the potential $V(T)$
within the observable scales. Since we approximate the power spectrum to the first order of the slow-roll
parameters by assuming that the higher order corrections are small,
the reconstruction is valid only under the slow-roll approximation and the reconstructed potential
satisfies the slow-roll condition. Outside the slow-roll regime, the potential can be rather different.

\subsection{The relation to generalized $\beta$-function formalism}

In the generalized $\beta$-function formalism \cite{Binetruy:2014zya,Pieroni:2015cma,Binetruy:2016hna}, the superpotential $W(T)=-2H(T)$ and the $\beta$-function is defined as
\begin{equation}
\label{betaeq1}
\beta(T)=-2(-p_{,X})^{-1/2}\frac{W_{,T}}{W},
\end{equation}
where $-p_{,X}=\rho=V(T)/\sqrt{1-\dot T^2}$. By using the Friedman eqs. \eqref{frweq1} and \eqref{frweq2},
it can be shown that $\beta^2(T)=3\dot T^2$, so
\begin{equation}
\label{betaeq2}
\beta(T)=-\sqrt{2\epsilon}\approx -\frac{V_{,T}}{V^{3/2}}.
\end{equation}
For a given $\beta$-function, we can reconstruct the potential $V(T)$ from the above relation.
Once the potential is reconstructed, we can derive the parametrization from eq. \eqref{betaeq2}.
Alternatively, if we parameterize the slow-roll parameters or the observables by $N$, we can
reconstruct the potential and derive the $\beta$-function from eq. \eqref{betaeq2}.

\section{The reconstruction}

The simplest parametrization is the constant parametrization. Let us consider $\epsilon=r/16$ being a constant first.
For this case, we get $\eta=0$ from eq. \eqref{slreq10}, and $n_s=1-2\epsilon=1-r/8$ from eq. \eqref{nseq2}.
The result $r=8(1-n_s)$ is excluded by the Planck 2015 observations at the $3\sigma$ level.
If we assume that $\eta$ is a constant, then from eq. \eqref{slreq10}, we get
\begin{equation}
\label{slrsol1}
\epsilon=e^{-\eta N},
\end{equation}
where we choose the integration constant so that $\epsilon(N=0)=1$. Plugging the result \eqref{slrsol1} into eq. \eqref{nseq2},
we get
\begin{equation}
\label{nssol1}
n_s=1-\frac{r}{8}+\frac{1}{N}\ln\left(\frac{r}{16}\right).
\end{equation}
The result is also excluded by the Planck 2015 observations at the $3\sigma$ level.
Now let us consider the case that $n_s$ is a constant.
From eq. \eqref{nseq2}, we get
\begin{gather}
\label{nssol2}
\epsilon=-\frac{\alpha e^{\alpha N}}{2e^{\alpha N}+\alpha D},\\
\label{nssol3}
V=-V_0 \left[1+\frac{2}{\alpha D}e^{\alpha N}\right]^{-1},
\end{gather}
where the constant $\alpha=n_s-1$. The end of inflation $\epsilon(N=0)=1$ gives $D=-(2+\alpha)/\alpha$.
The tensor to scalar ratio is
\begin{equation}
\label{nssol4}
r=\frac{16(1-n_s)}{2-(1+n_s)e^{(1-n_s)N}}.
\end{equation}
This result is again excluded by the Planck 2015 observations at the $3\sigma$ level.

\subsection{The constant slow-roll inflation}

\begin{figure}
$\begin{array}{cc}
\includegraphics[width=0.4\textwidth]{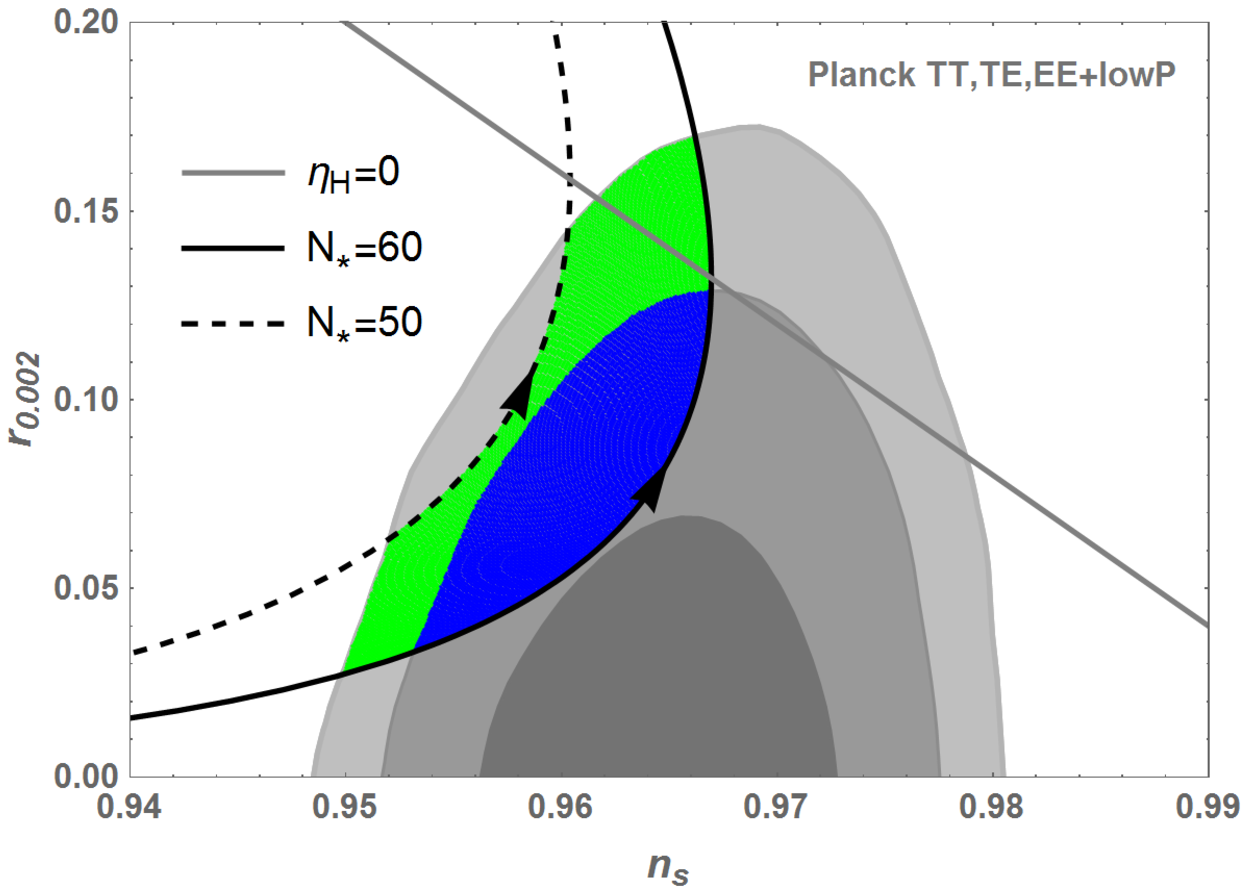}&
\includegraphics[width=0.4\textwidth]{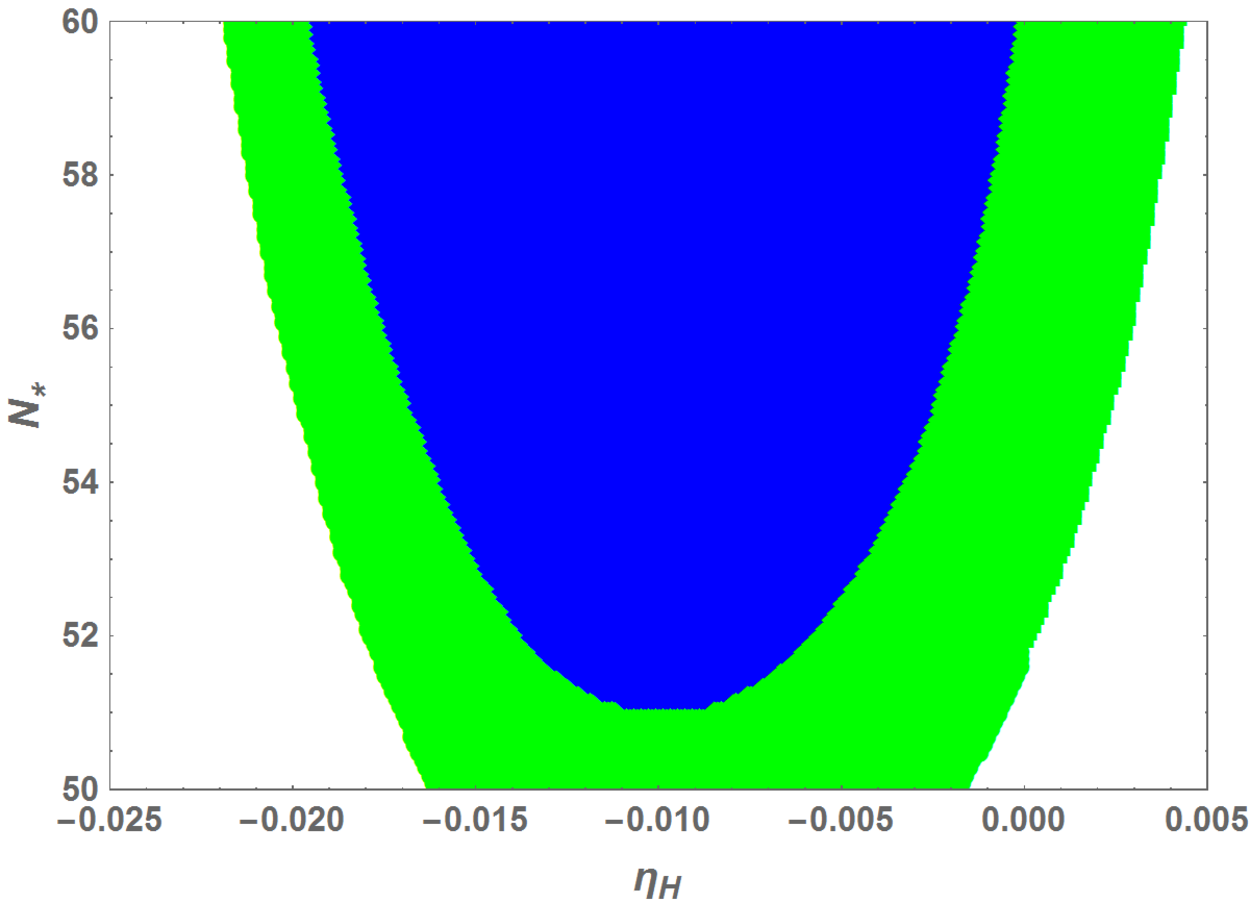}
\end{array}$
\caption{The marginalized 68\%, 95\% and 99.8\% confidence level contours for $n_s$ and $r_{0.002}$ from Planck 2015 data \citep{Ade:2015lrj} and the observational
constraint on $\eta_H$.
The left panel shows the $n_s-r$ contours and $\eta_H$ increases along the arrow direction.
The right panel shows the 95\% and 99.8\% confidence level constraints on $\eta_H$ and $N_*$
and they are colored by the blue and green, respectively.}
\label{fig1h}
\end{figure}

Now Let us consider the slow-roll parametrization with constant $\eta_H$,
\begin{equation}
\label{etaheq1}
\eta_H=-\frac{\ddot H}{2H\dot{H}}=\epsilon+\frac{1}{2}\frac{d\ln\epsilon}{dN},
\end{equation}
where the constant $|\eta_H|<1$.
By imposing the condition $\epsilon(N=0)=1$, the solution to eq. \eqref{etaheq1} is
\begin{equation}
\label{etaheq2}
\epsilon=\frac{r}{16}=\frac{\eta_H}{1+(\eta_H-1) \exp(-2\eta_H N)}.
\end{equation}
Substituting the result \eqref{etaheq2} into eqs. \eqref{nseq2} and \eqref{tneq1}, we get
\begin{equation}
\label{etaheq3}
n_s-1=2\eta_H-\frac{4\eta_H}{1+(\eta_H-1) \exp(-2\eta_H N)},
\end{equation}
and the reconstructed potential
\begin{equation}
V=2V_0\sec^{2}\left[\sqrt{\eta_H V_0}\,(T-T_0)\right].
\end{equation}
Comparing the results \eqref{etaheq2} and \eqref{etaheq3} with the Planck 2015 observations \cite{Ade:2015lrj},
we get the constraints on $\eta_H$ and $N_*$ and the results are shown in figure \ref{fig1h}. We see that the constant $\eta_H$
is not consistent with the observations at the $1\sigma$ level if $N_*\le 60$.

\begin{figure}
$\begin{array}{cc}
\includegraphics[width=0.4\textwidth]{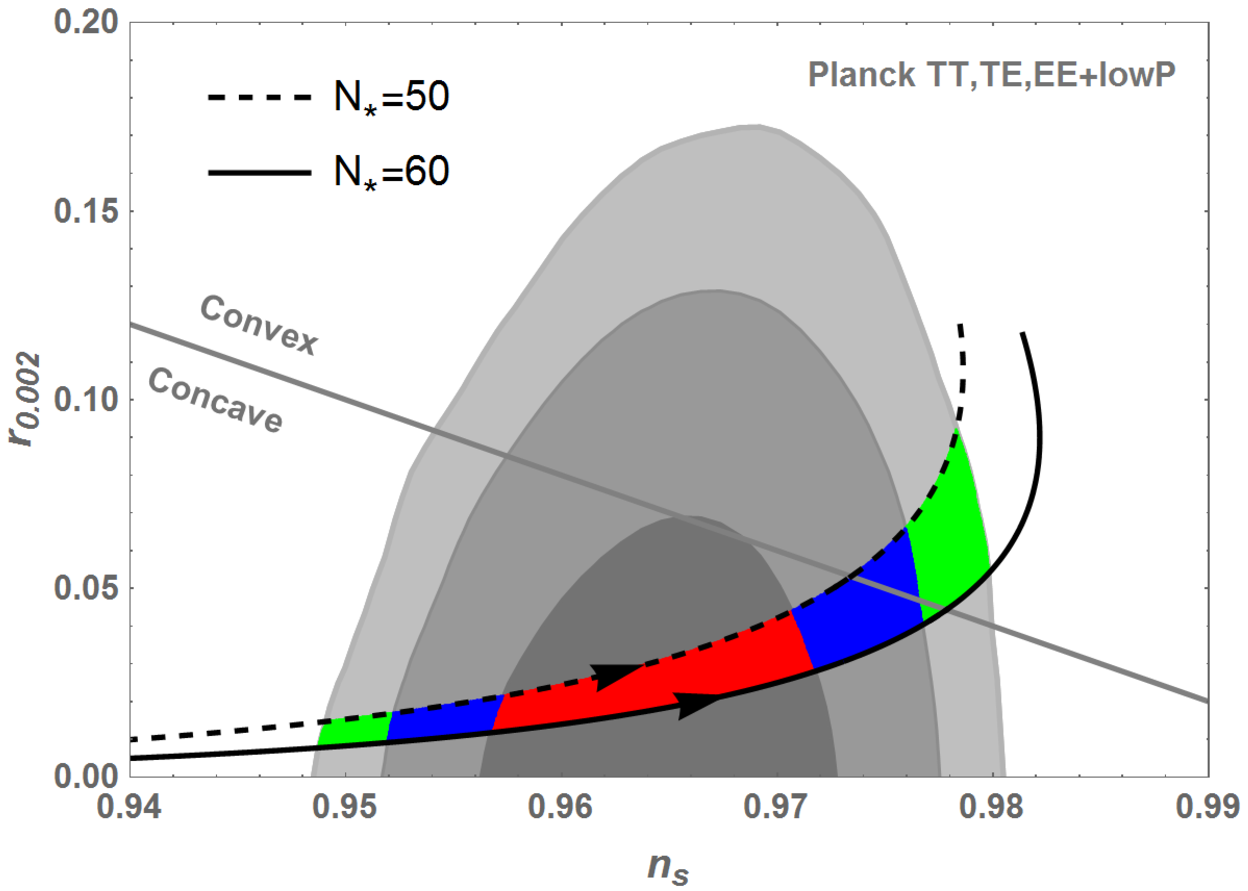}&
\includegraphics[width=0.4\textwidth]{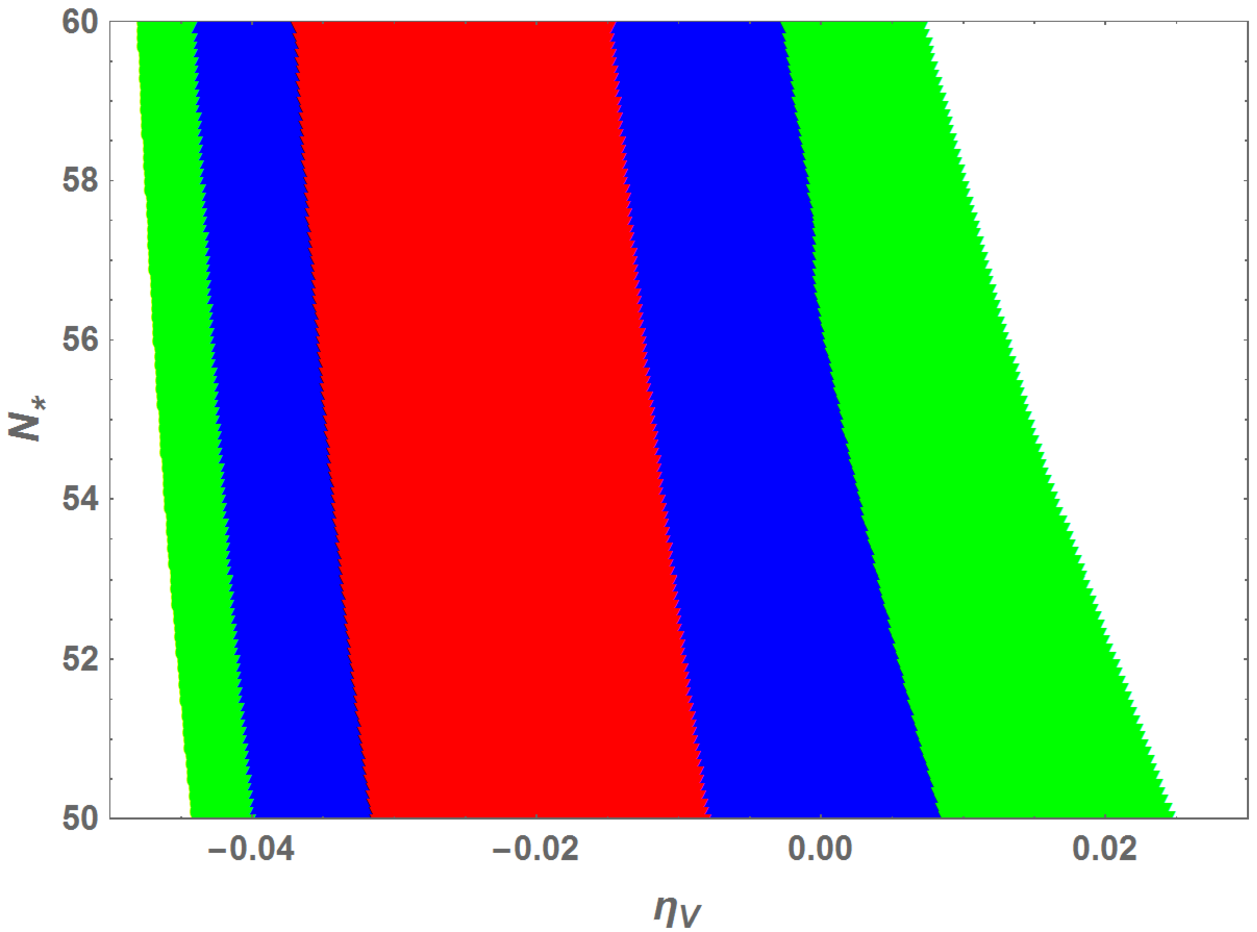}
\end{array}$
\caption{The marginalized 68\%, 95\% and 99.8\% confidence level contours for $n_s$ and $r_{0.002}$ from Planck 2015 data \citep{Ade:2015lrj} and the observational
constraint on $\eta_V$.
The left panel shows the $n_s-r$ contours and $\eta_V$ increases along the arrow direction.
The right panel shows the 68\%, 95\% and 99.8\% confidence level constraints on $\eta_V$ and $N_*$
and they are colored by the red, blue and green, respectively.}
\label{fig1}
\end{figure}

Next we consider the constant slow-roll parametrization
\begin{equation}
\label{conseq2}
\eta_V=2\frac{V_{,TT}}{V^2}\approx \frac{d\ln\epsilon}{dN}+6\epsilon,
\end{equation}
where the constant $|\eta_V|<1$. From the definition of $\eta_V$,  we find that the potential $V(T)$ takes the form of the Weierstrass function.
By imposing the condition $\epsilon(N=0)=1$, the solution to eq. \eqref{conseq2} is
\begin{equation}
\label{conseq3}
\epsilon=\frac{r}{16}=-\frac{\eta_V}{(6-\eta_V)e^{-\eta_V N}-6}.
\end{equation}
Substituting eq. \eqref{conseq3} into eq. \eqref{nseq2}, we get
\begin{equation}
\label{conseq4}
n_s-1=\frac{8\eta_V}{(6-\eta_V)e^{-\eta_V N}-6}+\eta_V.
\end{equation}
From eqs. \eqref{conseq3} and \eqref{conseq4}, we get $\eta_V=n_s-1+r/2$.
Comparing the results \eqref{conseq3} and \eqref{conseq4} with the Planck 2015 observations \cite{Ade:2015lrj},
we get the constraints on $\eta_V$ and $N_*$ and the results are shown in figure \ref{fig1}.
For $N_*=60$, we get $-0.0374< \eta_V< -0.0142$ at the $1\sigma$ level, $-0.0435 < \eta_V< -0.0031$ at the  $2\sigma$ level and $-0.0473 <\eta_V< 0.0067$
at the $3\sigma$ level.
From figure \ref{fig1}, we see that $\eta_V<0$ is favored at more than $2\sigma$ confidence level, so the concave potential is preferred.
From now on, we call constant $\eta_V$ as the constant slow-roll inflation.

\begin{figure}
\includegraphics[width=0.45\textwidth]{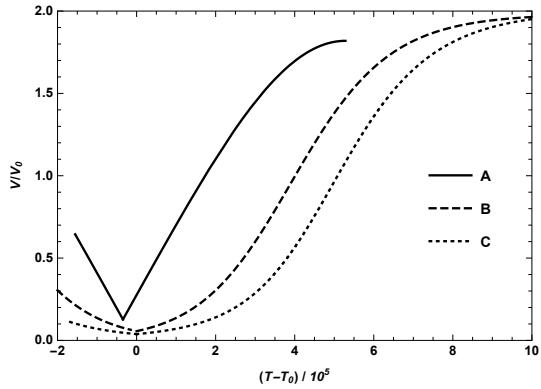}
\caption{The reconstructed potentials normalized by $V_0$ from eq. \eqref{conseq5a}. The solid line A corresponds to the potential for the constant slow-roll inflation,
the dashed line B denotes the potential \eqref{nspareq9}, and the dotted line C denotes the potential from eq. \eqref{rpareq10}.}
\label{fig1v}
\end{figure}

\begin{figure}
\includegraphics[width=0.45\textwidth]{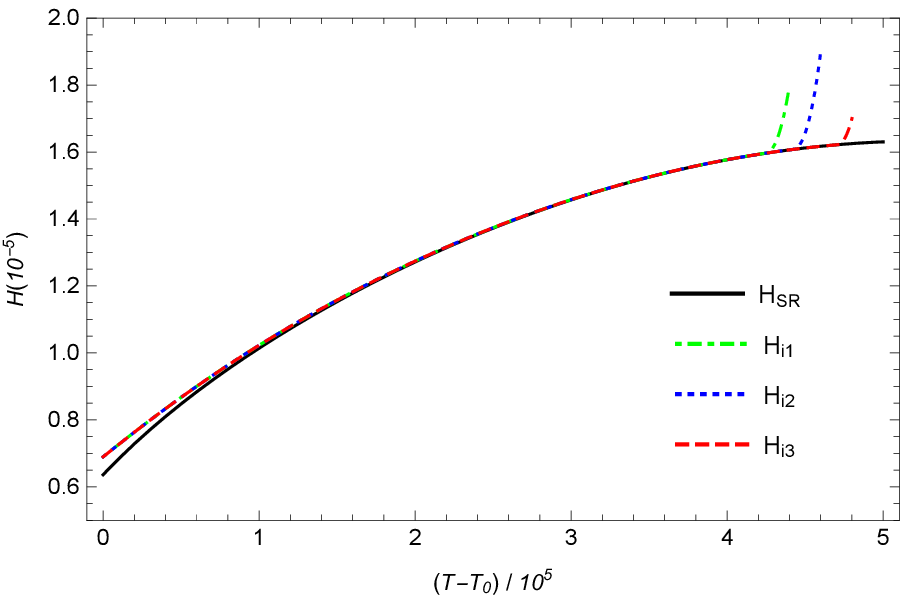}
\caption{The solutions to the Hamilton-Jacobi equation \eqref{H-J} for the potential \eqref{conseq5} with different initial values $H(T_i)$.
The solid line labelled as $H_{\text{SR}}$ corresponds to the slow-roll attractor.}
\label{figatta}
\end{figure}

Substituting eq. \eqref{conseq3} into eq. \eqref{slreq9}, we get
\begin{equation}
\label{conseq5}
V(N)=V_0\left|6e^{\eta_V N}-6+\eta_V\right|^{\frac 13},
\end{equation}
where
\begin{equation}
\label{conseq5a}
V_0=\frac{3\pi^2 A_s r}{2}\left|\frac{8-8n_s-r}{\left(8-8n_s-4r\right)\left(7-n_s-r/2\right)}\right|^{1/3}.
\end{equation}
From eq. \eqref{tneq1}, we get
\begin{equation}
\label{conseq6}
d T=\frac{2}{\eta_V}\sqrt{\frac{2|\eta_V|}{ V_0}}\,\frac{1}{(6x^2-6+\eta_V)^{2/3}}dx,
\end{equation}
where $x=e^{\eta_V N/2}$. The solution gives the Hypergeometric function
\begin{equation}
\label{conseq7}
T-T_0=\frac{2}{\eta_V}\sqrt{\frac{2|\eta_V|}{ V_0}}\frac{\exp(\eta_V N/2)}{(6-\eta_V)^{2/3}}\, _2F_1\left(\frac{1}{2},\frac{2}{3};\frac{3}{2};\frac{6e^{\eta_V N}}{6-\eta_V}\right).
\end{equation}
Combining eqs. \eqref{conseq5} and \eqref{conseq7}, we can obtain the potential $V(T)$.
If we take $\eta_V=-0.021$ and $N_*=60$, we get $n_s=0.968$, $r=0.022$ and $\Delta T=T_*-T_e=3.66\times 10^5$, so the field excursion satisfies the bound \eqref{gfbld2}.
By using these parameters, we plot the potential in figure \ref{fig1v}, and the slow-roll attractor for the potential is shown in figure \ref{figatta}.

\subsection{The power-law parametrization of $n_s$}

The observational data favors $n_s=1-2/N*$ with $N_*=60$, so we choose the parametrization
\begin{equation}
\label{nspareq5}
n_s=1-\frac{p}{N+A}.
\end{equation}
From eq. \eqref{nseq2}, we get
\begin{equation}
\label{nspareq1}
\epsilon(N)=\begin{cases}
\frac{\displaystyle p-1}{\displaystyle 2(N+A)+2C(p-1)(N+A)^p},& p\neq1,\\
\frac{\displaystyle 1}{\displaystyle 2(N+A)[C-\ln(N+A)]},& p=1,
\end{cases}
\end{equation}
and
\begin{equation}
\label{nspareq2}
V(N)=\begin{cases}
\frac{\displaystyle \tilde V_0(p-1)}{\displaystyle (p-1)C+(N+A)^{1-p}},& p\neq 1,\\
\frac{\displaystyle \tilde V_0}{\displaystyle C-\ln(N+A)}, & p=1,
\end{cases}
\end{equation}
where $C$ and $\tilde V_0$ are integration constants.

For convenience, let us consider the case $p=1$ first.
At the end of inflation, $N=0$ and $\epsilon(N)=1$, so $C=(2A)^{-1}+\ln A$, and we get
\begin{gather}
\label{nspareq3}
n_s=1-\frac{1}{N+A},\\
\label{nspareq4}
r=\frac{16A}{(N+A)\left(1-A\ln [(N+A)/A]^2\right)}.
\end{gather}
It is easy to show that the results are excluded by the Planck 2015 observations \cite{Ade:2015lrj} at the $3\sigma$ level.

\begin{figure}
$\begin{array}{cc}
\includegraphics[width=0.4\textwidth]{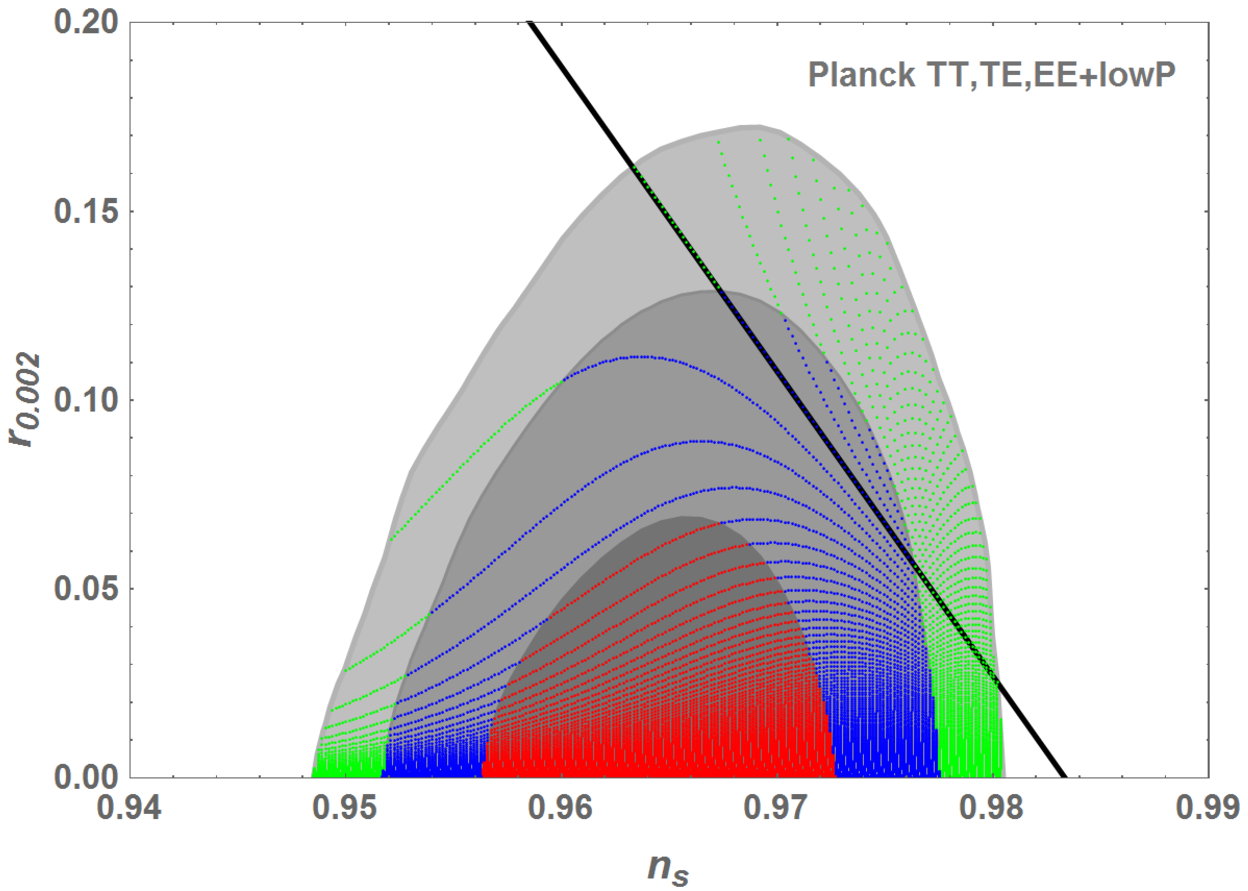}&
\includegraphics[width=0.4\textwidth]{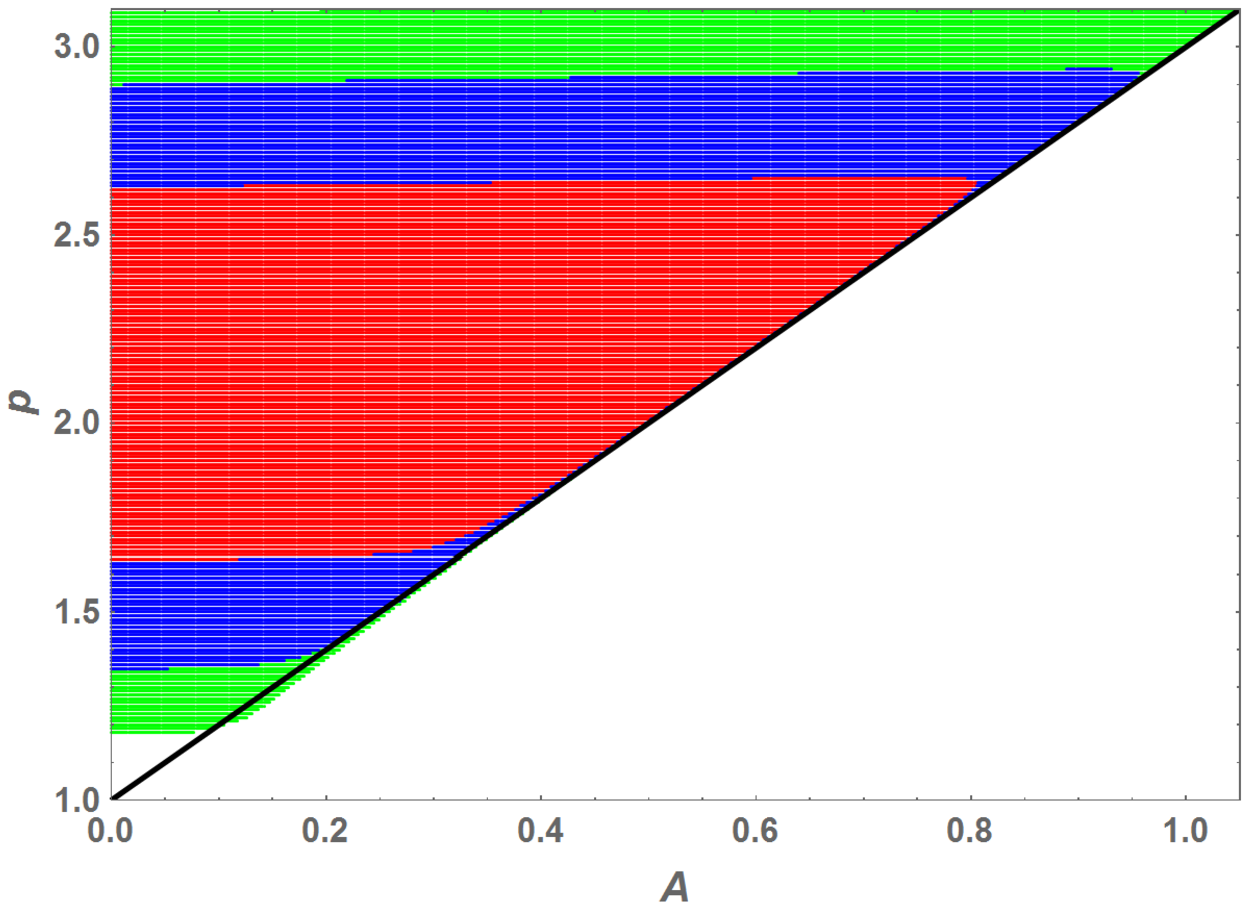}
\end{array}$
\caption{The marginalized 68\%, 95\% and 99.8\% confidence level contours for $n_s$ and $r_{0.002}$ from Planck 2015 data \citep{Ade:2015lrj}
and the theoretical predictions for the parametrization \eqref{nspareq5} with $N_*=60$.
The left panel shows the $n_s-r$ contours and the right panel shows the constraints on $p$ and $A$ for $N_*=60$.
The red, blue and green regions correspond to 68\%, 95\% and 99.8\% confidence levels, respectively. The solid black line denotes $p=1+2A$.}
\label{fig2}
\end{figure}

For $p\neq 1$, from the condition $\epsilon(N=0)=1$, we get $p-1-2A=2C(p-1)A^p$, so the tensor to scalar ratio is
\begin{equation}
\label{nspareq6}
r=\frac{16(p-1)}{2(N+A)+(p-1-2A)(N+A)^p/A^p}.
\end{equation}
Note that if $C=0$, then $p=1+2A$ and $r=8[N(1-n_s)-1]/(N-1/2)$. If $p>1+2A$ with $A>0$, then $r\sim 1/N^p$.
In particular, for the case $p=2$, we get the familiar $\alpha$ attractor
$n_s=1-2/N_*$ and $r=12\alpha/N_*^2$ for large $N$ and $C\neq 0$.
Comparing the results \eqref{nspareq5} and \eqref{nspareq6} with the Planck 2015 observations \cite{Ade:2015lrj},
taking $N_*=60$,
we get the constraints on $p$ and $A$ and the results are shown in figure \ref{fig2}. From figure \ref{fig2}, we see
that the results are similar to those for canonical scalar field \cite{Lin:2015fqa}.

\begin{figure}
\includegraphics[width=0.45\textwidth]{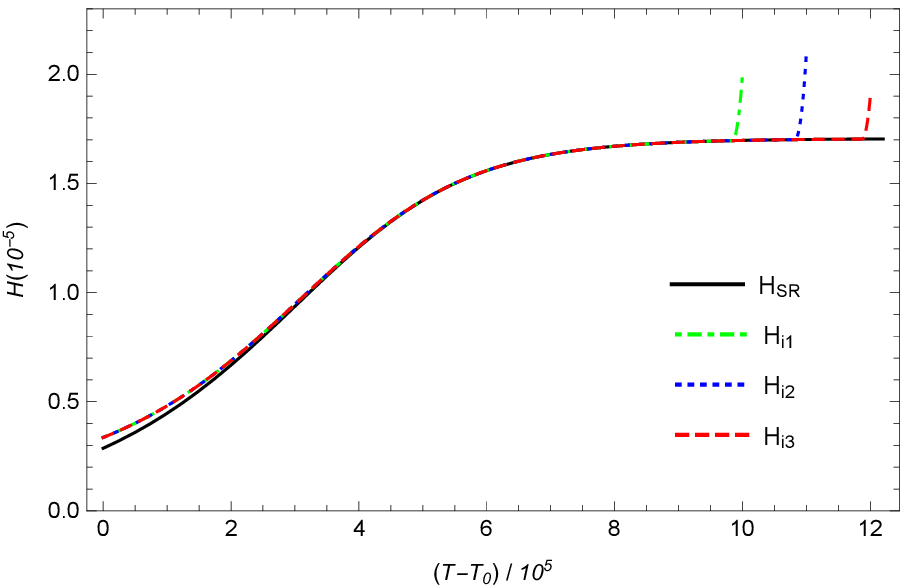}
\caption{The solutions to the Hamilton-Jacobi equation \eqref{H-J} for the potential \eqref{nspareq9} with different initial values $H(T_i)$.
The solid line labelled as $H_{\text{SR}}$ corresponds to the slow-roll attractor.}
\label{figattb}
\end{figure}

Now we proceed to derive the class of potentials.
From eq. \eqref{tneq1}, we get
\begin{equation}
\label{nspareq7}
dT=\tilde V_0^{-1/2}(N+A)^{-\frac p2}dN.
\end{equation}
For $p\neq 1$ and $p\neq 2$, we have
\begin{equation}
\label{nspareq8}
  N(T)+A=\left[\frac{(2-p)\sqrt{\tilde V_0}}{2}(T-T_0)\right]^{\frac2{2-p}}.
\end{equation}
Combining eqs. \eqref{nspareq2} and \eqref{nspareq8}, for $p\neq 1+2A$, $p\neq 1$ and $p\neq 2$,
we get the inverse power-law potential \cite{Steer:2003yu,Brax:2003rs,Abramo:2003cp}
\begin{equation}
\label{nspareq9}
V(T)=V_0\left[1+\beta_1 (T-T_0)^{\frac{2p-2}{p-2}}\right]^{-1},
\end{equation}
where
\begin{gather}
\label{nspareq10}
V_0=\frac{2\tilde V_0(p-1)A^p}{p-1-2A},\\
\label{nspareq11}
\beta_1=\frac{2A^p\tilde V_0^{(p-1)/(p-2)}}{p-1-2A}\left(\frac{2-p}{2}\right)^{2(p-1)/(p-2)},
\end{gather}
and
\begin{equation}
\label{nspareq16}
V(T)=(p-1)\tilde{V}_0\left[\frac{(2-p)\sqrt{\tilde{V_0}}}{2}(T-T_0)\right]^{2(p-1)/(2-p)},
\end{equation}
for $p=1+2A$, $p\neq 1$ and $p\neq 2$.
If we take $p=1.934$, $A=0.446$ and $N_*=60$, we get $n_s=0.968$, $r=0.022$, and
$\Delta T=T_*-T_e=5.4\times 10^5$, so the bound \eqref{gfbld2} is satisfied.
With these model parameters,
we plot the potential \eqref{nspareq9} in figure \ref{fig1v}, and
the slow-roll attractor is shown in figure \ref{figattb}.

For $p=2$, we have
\begin{equation}
\label{nspareq12}
N(T)+A=\exp\left[\sqrt{\tilde V_0}\,(T-T_0)\right].
\end{equation}
Combining eqs. \eqref{nspareq2} and \eqref{nspareq12}, we get the potential for the $\alpha$ attractor \eqref{nspareq5} with $p=2$
\begin{equation}
\label{nspareq13}
V(T)=V_0\left[1+\beta_2 \exp\left(-\sqrt{\tilde V_0}\,(T-T_0)\right)\right]^{-1},
\end{equation}
where $A\neq 1/2$ and
\begin{gather}
\label{nspareq14}
V_0=\frac{2A^2\tilde V_0}{1-2A},\\
\label{nspareq15}
\beta_2=\frac{2A^2}{1-2A},
\end{gather}
and
\begin{equation}
\label{nspareq17}
V(T)=\tilde{V}_0\exp\left[\sqrt{\tilde{V}_0}\,(T-T_0)\right],
\end{equation}
for $A=1/2$ and $p=2$.

 \subsection{The power-law parametrization of $r$}

\begin{figure}
$\begin{array}{cc}
\includegraphics[width=0.4\textwidth]{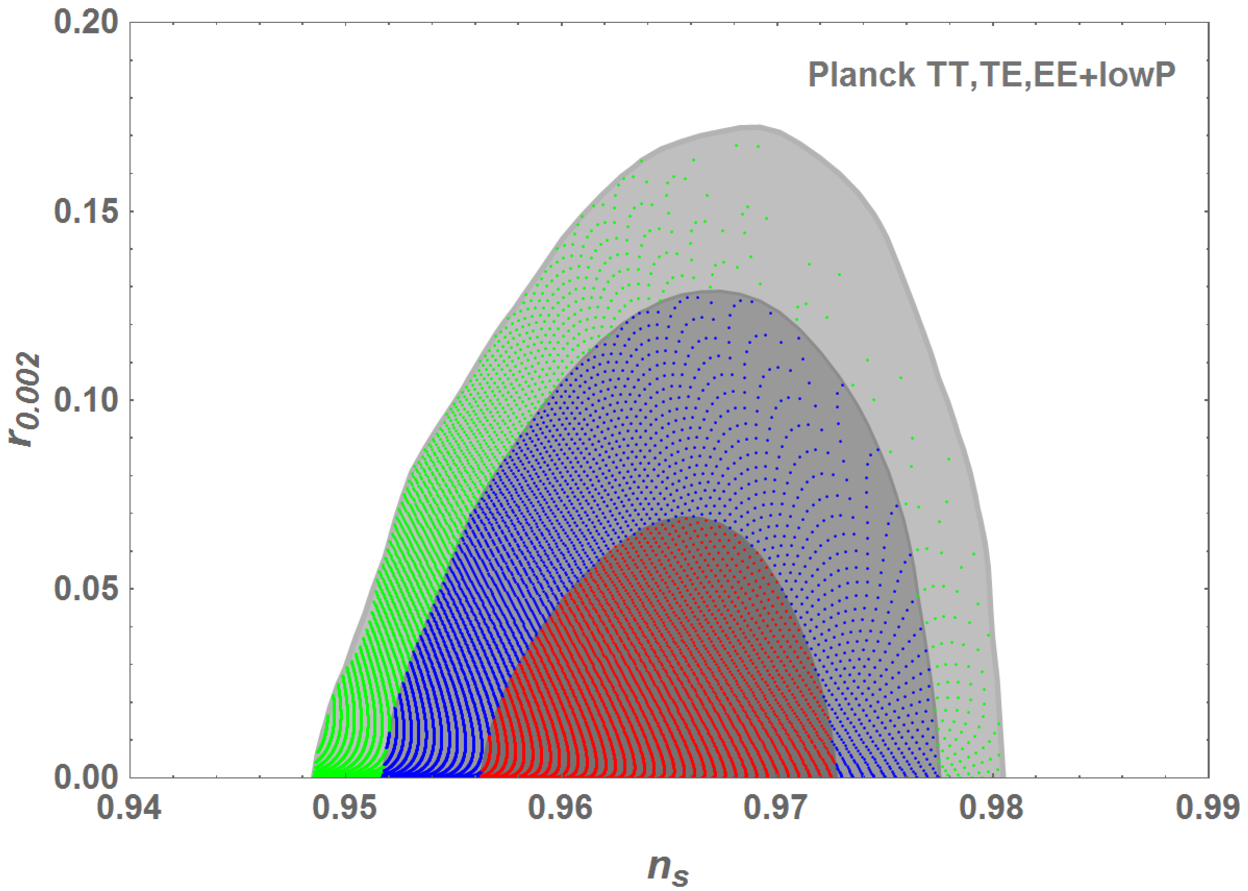}&
\includegraphics[width=0.4\textwidth]{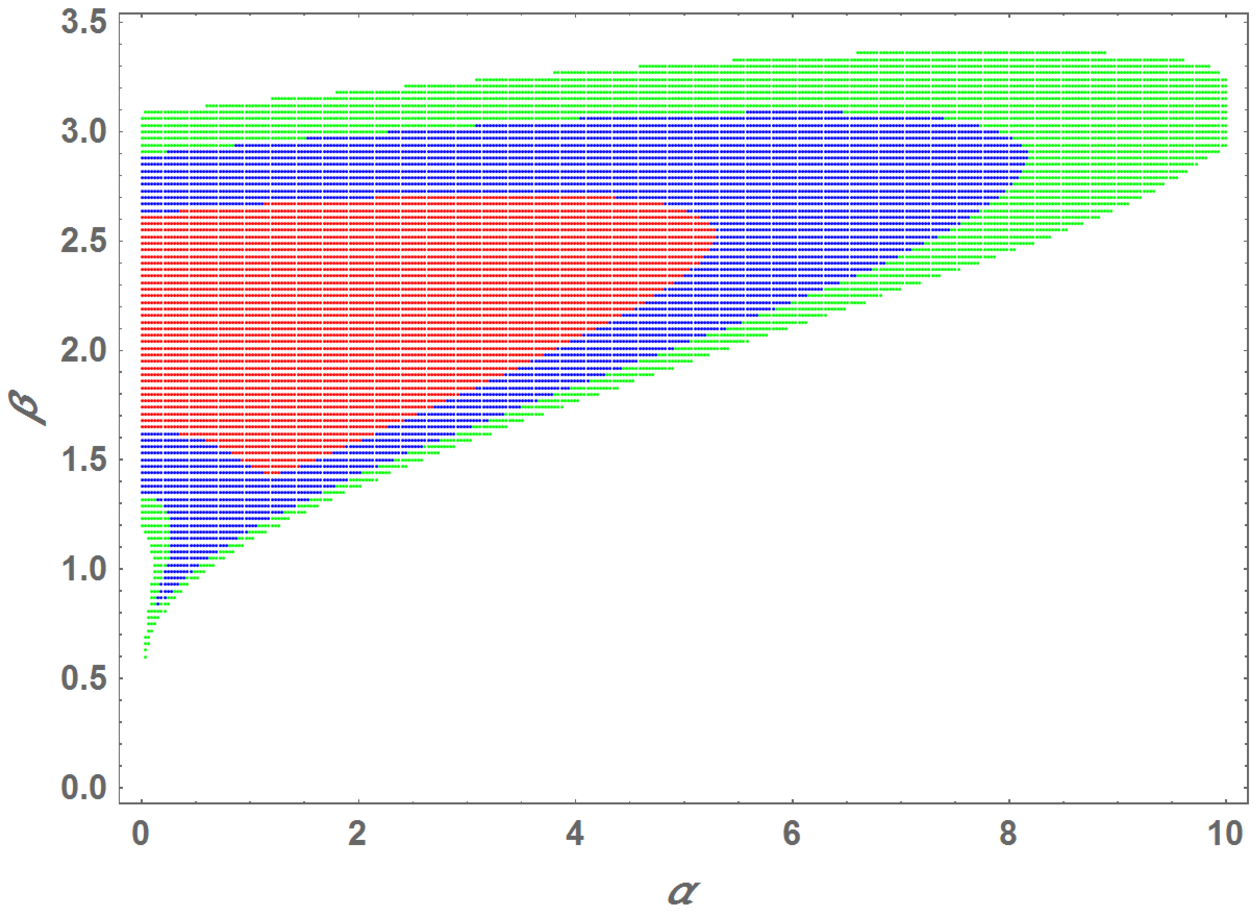}
\end{array}$
\caption{Same as figure \ref{fig2} but for the parametrization \eqref{rpareq1}.
The left panel shows the $n_s-r$ contours and the right panel shows the constraints on $\alpha$ and $\beta$ for $N_*=60$.
The red, blue and green regions correspond to 68\%, 95\% and 99.8\% confidence levels, respectively.}
\label{fig3}
\end{figure}

In this subsection, we consider the parametrization
\begin{equation}
\label{rpareq1}
\epsilon=\frac{r}{16}=\frac \gamma{(N+\alpha)^\beta},
\end{equation}
where $\gamma=\alpha^\beta$ so that $\epsilon(N=0)=1$. Substituting the parametrization \eqref{rpareq1} into eq. \eqref{nseq2}, we get
\begin{equation}
\label{rpareq2}
n_s-1=-2(\frac{\alpha}{N+\alpha})^\beta-\frac{\beta}{N+\alpha}.
\end{equation}
Comparing the results \eqref{rpareq1} and \eqref{rpareq2} with the Planck 2015 observations \cite{Ade:2015lrj},
we get the constraints on $\alpha$ and $\beta$ and the results are shown in figure \ref{fig3}.
Substituting eq. \eqref{rpareq1} into eq. \eqref{slreq9}, we get
\begin{equation}
\label{rpareq3}
V(N)=\tilde{V}_0(N+\alpha)^{2\alpha},
\end{equation}
for $\beta=1$, and
\begin{equation}
\label{rpareq4}
V=\tilde{V}_0\exp\left[\frac{2\gamma}{1-\beta}(N+\alpha)^{1-\beta}\right],
\end{equation}
for $\beta\neq 1$. So for $\beta\neq 1$, combining eqs. \eqref{rpareq1} and \eqref{rpareq4}, we get
\begin{equation}
\label{rpareq4a}
\epsilon=\gamma\left[\frac{1-\beta}{2\gamma}\ln\left(\frac{V}{\tilde{V}_0}\right)\right]^{-\frac{\beta}{1-\beta}}.
\end{equation}

Let us consider the case $\beta=1$ first. Substituting eq. \eqref{rpareq3} into eq. \eqref{tneq1}, we get
\begin{equation}
\label{rpareq5}
\frac{d T}{dN}=\sqrt{\frac{2\alpha}{\tilde{V}_0}}\,(N+\alpha)^{-\alpha-\frac{1}{2}}.
\end{equation}

If $\alpha=1/2$, then
\begin{equation}
\label{rpareq6}
N+\alpha=\exp\left[\sqrt{\tilde{V}_0}\,(T-T_0)\right].
\end{equation}
Combining eq. \eqref{rpareq3} and eq. \eqref{rpareq6}, we get the exponential potential \cite{Barbosa-Cendejas:2015rba} for $\beta=1$ and $\alpha=1/2$
\begin{equation}
\label{rpareq7}
V=\tilde{V}_0\exp\left[\sqrt{\tilde{V}_0}\,(T-T_0)\right].
\end{equation}

If $\alpha \neq 1/2$, then
\begin{equation}
\label{rpareq8}
N+\alpha=\left[\sqrt{\frac{\tilde{V_0}}{2\alpha}}\,\left(\frac{1}{2}-\alpha\right)(T-T_0)\right]^{\frac{2}{1-2\alpha}}.
\end{equation}
Combining eq. \eqref{rpareq3} and eq. \eqref{rpareq8}, we get the power-law potential \cite{Barbosa-Cendejas:2015rba} for $\beta=1$ and $\alpha\neq 1/2$
\begin{equation}
\label{rpareq9}
V=V_0(T-T_0)^{4\alpha/(1-2\alpha)},
\end{equation}
where
\begin{equation}
\label{rpareq9a}
V_0=\tilde{V}_0\left[\sqrt{\frac{\tilde{V_0}}{2\alpha}}\,\left(\frac{1}{2}-\alpha\right)\right]^{\frac{4\alpha}{1-2\alpha}}.
\end{equation}

\begin{figure}
\includegraphics[width=0.45\textwidth]{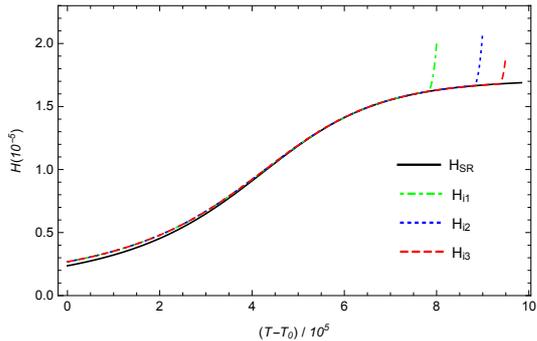}
\caption{The solutions to the Hamilton-Jacobi equation \eqref{H-J} for the potential \eqref{rpareq10} with different initial values $H(T_i)$.
The solid line labelled as $H_{\text{SR}}$ corresponds to the slow-roll attractor.}
\label{figattc}
\end{figure}

Now let us discuss the general case $\beta\neq 1$. Substituting eq. \eqref{rpareq4a} into eq. \eqref{tneq2}, we get
\begin{equation}
\label{rpareq10}
\sqrt{2\gamma}\, dT=\pm \left[\frac{1-\beta}{2\gamma}\ln\left(\frac{V}{\tilde{V}_0}\right)\right]^{\frac{\beta}{2(1-\beta)}} \frac{dV}{\sqrt{V^3}}.
\end{equation}
Although the analytic form for $V(T)$ is not available, the potential $V(T)$ can be obtained from  eq. \eqref{rpareq10}
and it is shown in figure \ref{fig1v}. The slow-roll attractor is shown in figure \ref{figattc}.
If we take $\alpha=1.588$, $\beta=1.801$ and $N_*=60$, we get $n_s=0.968$, $r=0.022$ and $\Delta T=T_*-T_e=6.93\times 10^5$, so the bound \eqref{gfbld2} is satisfied.

\section{Reheating}

From figure \ref{fig1v}, we see that the reconstructed potentials have minimum, so when the tachyon rolls down to the minimum,
inflation ends and the tachyon field begins to oscillate around the minimum.
Due to the interaction between the tachyon and relativistic particles, the tachyon decays to relativistic particles
and the energy stored in the tachyon field is converted to relativistic particles. Although the physics of reheating is uncertain,
the reheating process may provide additional constraint on inflationary models.

The pivotal scale $k_\ast=0.002\text{Mpc}^{-1}$ is related to the current Hubble horizon as
\begin{equation}
\label{kstar}
  \frac{c_sk_\ast}{a_0 H_0}=\frac{a_\ast H_\ast}{a_0 H_0}=\frac{a_\ast}{a_{e}}\frac{a_{e}}{a_{re}}\frac{a_{re}}{a_0}\frac{H_\ast}{H_0}=e^{-N_\ast-N_{re}}\frac{a_{re}}{a_0}\frac{H_\ast}{H_0},
\end{equation}
where $a_{re}$ denotes the value of the scale factor at the end of reheating, $N_{re}$ denotes the number of $e$-folds during reheating,
and we assume that radiation domination begins immediately after the reheating, and reheating begins immediately after inflation.
If the equation of state parameter $w_{re}$ is a constant during reheating, then we have
\begin{equation}
\label{reheq2}
N_{re}=\frac{1}{3(1+w_{re})}\ln\frac{\rho_{e}}{\rho_{re}},
\end{equation}
where $\rho_{re}$  is related with the temperature $T_{re}$ as
\begin{equation}
\label{reheq1}
\rho_{re}=\frac{\pi^2}{30}g_{re}T^4_{re},
\end{equation}
and $g_{re}$ is the effective number of relativistic species at reheating. From the entropy conservation, we can express the temperature $T_{re}$
with the current cosmic microwave background temperature $T_0=2.725K$ through the following relation
\begin{equation}
\label{reheq3}
a_{re}^3 g_{s,re}T_{re}^3=a_0^3\left(2 T_0^3+6\times\frac78 T_{\nu0}^3\right),
\end{equation}
where $g_{s,re}$ is the effective number of relativistic species for entropy and the current neutrino temperature $T_{\nu0}=(4/11)^{1/3}T_0$.
Combining the above results, we get \cite{Dai:2014jja,Cook:2015vqa}
\begin{gather}
\label{Nre}
N_{re}=\frac{4}{1-3w_{re}}\left[-N_\ast-\ln\frac{\rho_{e}^{1/4}}{H_\ast}+\frac{1}{3}\ln\frac{43}{11g_{s,re}}+
  \frac14\ln\frac{\pi^2 g_{re}}{30}-\ln\frac{c_sk_{\ast}}{a_0T_0}\right],\\
\label{Tre}
T_{re}=\exp\left[-\frac{3N_{re}(1+w_{re})}{4}\right]\left[\frac{30\rho_{e}}{\pi^2 g_{re}}\right]^{1/4}.
\end{gather}
Since $N_{re}$ and $T_{re}$ depend on $g_{re}$ and $g_{s,re}$ logarithmically, so it is safe to take $g_{re}=g_{s,re}=106.75$.
Since at the end of inflation, $\dot T^2=2/3$, so $\rho_{e}=\sqrt{3}V_{e}$. Using the observational value \cite{Ade:2015lrj}
 \begin{equation}
 \label{reheq4}
   A_s=H_{\ast}^2/(8\pi^2 \epsilon_{\ast})=2.2\times10^{-9},
 \end{equation}
we get
\begin{gather}
\label{Nre:final}
N_{re}=\frac{4}{1-3w_{re}}\left(56.94-N_\ast-\frac14\ln V_{e}+\frac12\ln\epsilon_\ast\right),\\
\label{reheq5}
T_{re}=\exp\left[-\frac{3N_{re}(1+w_{re})}{4}\right]\left[\frac{3\sqrt{3}V_{e}}{10.675 \pi^2}\right]^{1/4}.
\end{gather}
These results \eqref{Nre:final} and \eqref{reheq5} can be used to constrain inflationary models.

\begin{figure}
$\begin{array}{ccc}
\includegraphics[width=0.3\textwidth]{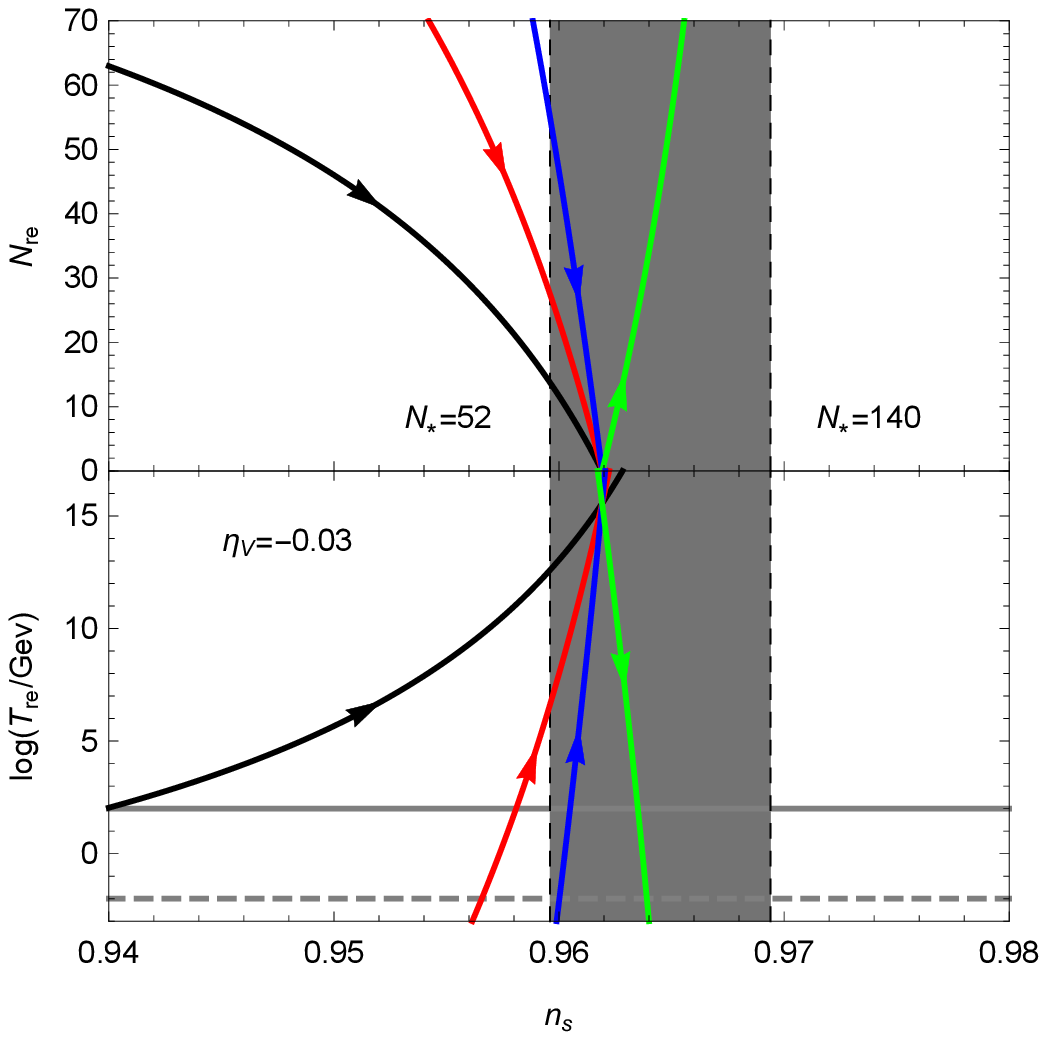}&
\includegraphics[width=0.3\textwidth]{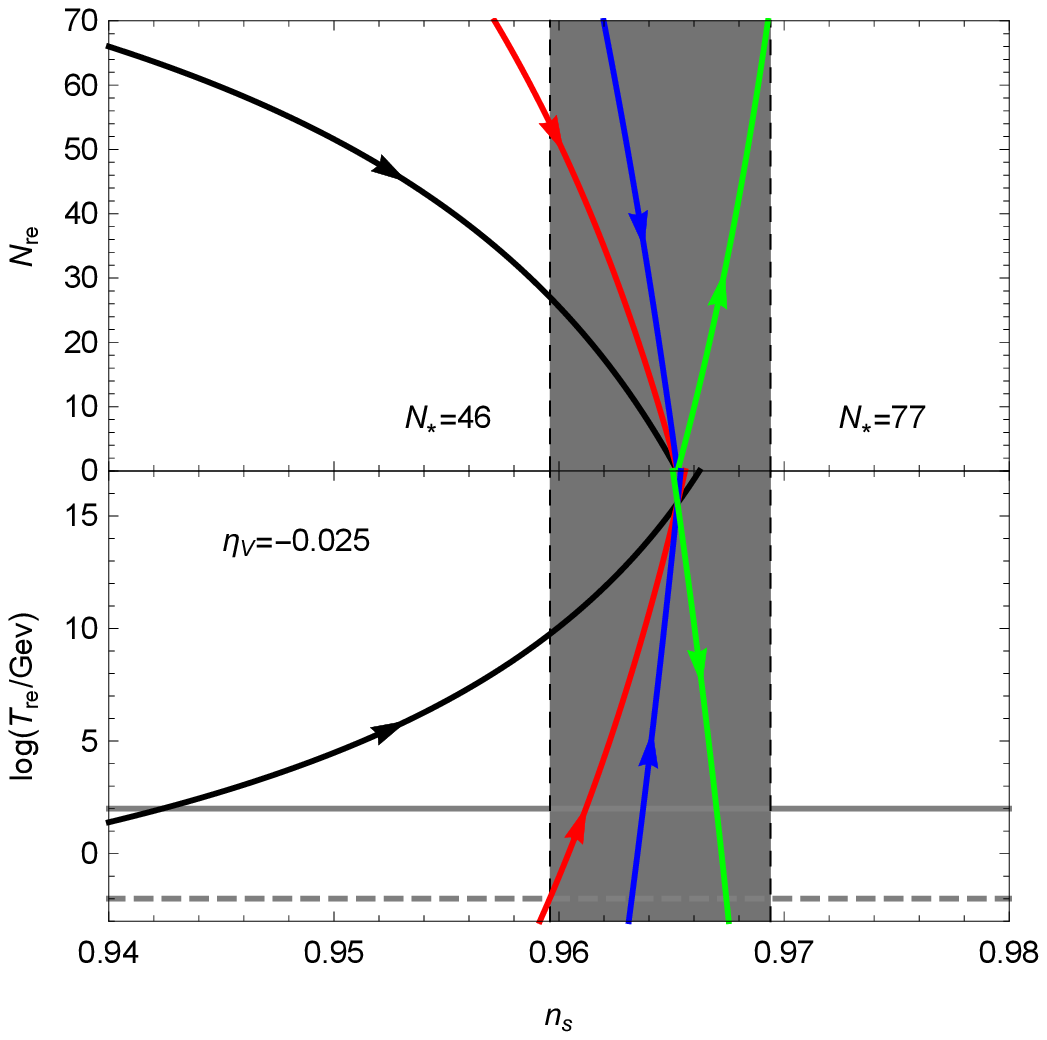}&
\includegraphics[width=0.3\textwidth]{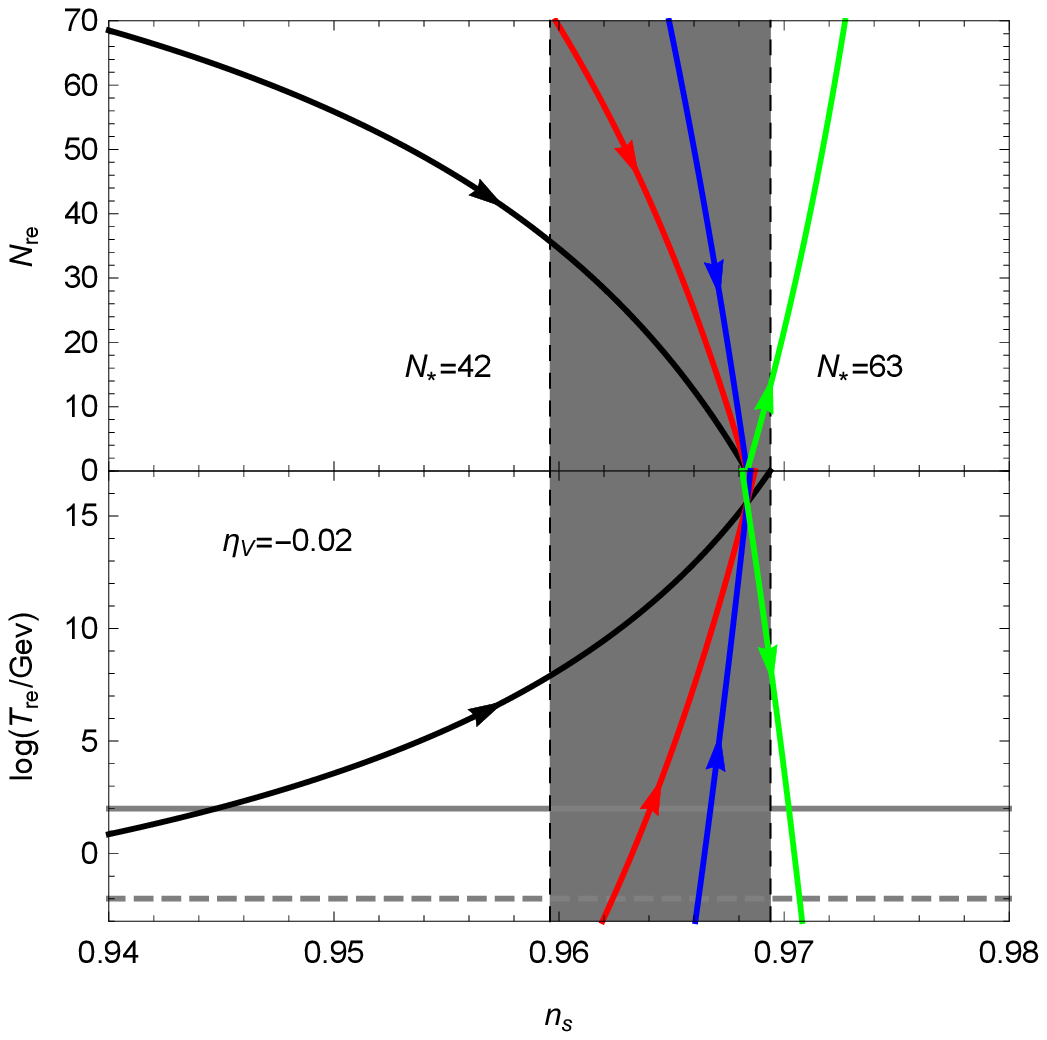}
\end{array}$
\caption{$N_{re}$ (upper panels) versus $n_s$ as determined from \eqref{conseq4} and \eqref{reheq8}, and $T_{re}$ (lower panels) versus $n_s$ as
determined from \eqref{conseq4} and \eqref{reheq9} for the constant slow-roll inflation.
From the left to right, the parameter $\eta_V$ is chosen as $\eta_V=-0.03$, $-0.025$ and $-0.02$, respectively.
The gray band corresponds to the $1\sigma$ Planck constraint $n_s=0.9645\pm 0.0049$ \cite{Ade:2015lrj},
and the $1\sigma$ constraint on $N_*$ is also given.
In each panel, the black, red, blue and green lines denote $w_{re}=-1/3$, 0, 1/6 and 2/3, respectively,
and the arrow indicates that $N_*$ increases along the line.
The horizontal gray solid and dashed lines in lower panels correspond to the electroweak scale $T_{EW}\sim 100$ Gev
and the big bang nucleosynthesis scale $T_{BBN}\sim 10$ Mev, respectively.}
\label{fig:reheating:eta}
\end{figure}

For the constant slow-roll inflation \eqref{conseq2}, at the horizon exit, we have
\begin{equation}
\label{reheq6}
3H_\ast^2=V_0\left|6e^{\eta_V N_\ast}-6+\eta_V\right|^{1/3}.
\end{equation}
At the end of inflation, $V_e=V_0|\eta_V|^{1/3}$, so
\begin{equation}
\label{reheq7}
V_{e}=24\pi^2\epsilon_* A_s \left[\frac{\eta_V e^{-\eta_V N_\ast}}{6-(6-\eta_V)e^{-\eta_V N_\ast}}\right]^{1/3}.
\end{equation}
In deriving the above result, we used the relation $H_*^2=8\pi^2\epsilon_* A_s$.
Substituting eqs. \eqref{conseq3} and \eqref{reheq7} into eqs. \eqref{Nre:final} and \eqref{reheq5}, we get
\begin{gather}
\label{reheq8}
N_{re}=\frac{4}{1-3w_{re}}\left\{60.56-\left(1-\frac{\eta_V}{12}\right)N_\ast+
  \frac{1}{6}\ln\left[\frac{\eta_V}{6-(6-\eta_V)e^{-\eta_V N_\ast}}\right]\right\},\\
\label{reheq9}
T_{re}=0.01\left[\frac{\eta_V}{6-(6-\eta_V)e^{-\eta_V N_\ast}}\right]^{1/3}\exp\left[-\frac{3N_{re}(1+w_{re})}{4}-\frac{\eta_V N_\ast}{12}\right].
\end{gather}
By choosing different values for $\eta_V$ and $w_{re}$, we calculate $n_s$, $N_{re}$ and $T_{re}$ by varying $N_*$, and the results are shown in figure \ref{fig:reheating:eta}.
The parameter $\eta_V$ is chosen as $\eta_V=-0.03$, $-0.025$ and $-0.02$ respectively from the left to right in figure \ref{fig:reheating:eta},
the gray region corresponds to the $1\sigma$ Planck constraint $n_s=0.9645\pm 0.0049$ \cite{Ade:2015lrj},
and the $1\sigma$ constraint on $N_*$ for the chosen value of $\eta_V$ is also shown.
The black, red, blue and green lines denote $w_{re}=-1/3$, 0, 1/6 and 2/3, respectively.
The horizontal gray solid and dashed lines in lower panels correspond to the electroweak scale $T_{EW}\sim 100$ Gev
and the big bang nucleosynthesis scale $T_{BBN}\sim 10$ Mev, respectively.
From figure \ref{fig:reheating:eta}, we see that depending on the model parameter $\eta_V$ and the
reheating physics (the value of $w_{re}$), the constraints on $N_{re}$ and $T_{re}$ are different.
As $\eta_V$ becomes larger, $n_s$ increases, the allowed reheating epoch becomes longer for $w_{re}=-1/3$, 0 and $1/6$
while the allowed reheating epoch becomes shorter for $w_{re}=2/3$.
For $-0.03<\eta_V<-0.02$, reheating with $-1/3\le w_{re}\le 2/3$ are all consistent with the observations.
Around the central value $n_s=0.965$, $\eta_V=-0.025$ and $N_*=60$, $w_{re}=1/6$ can have a prolonged reheating epoch and $N_{re}$ can be larger than 70.

\begin{figure}
$\begin{array}{cc}
\includegraphics[width=0.4\textwidth]{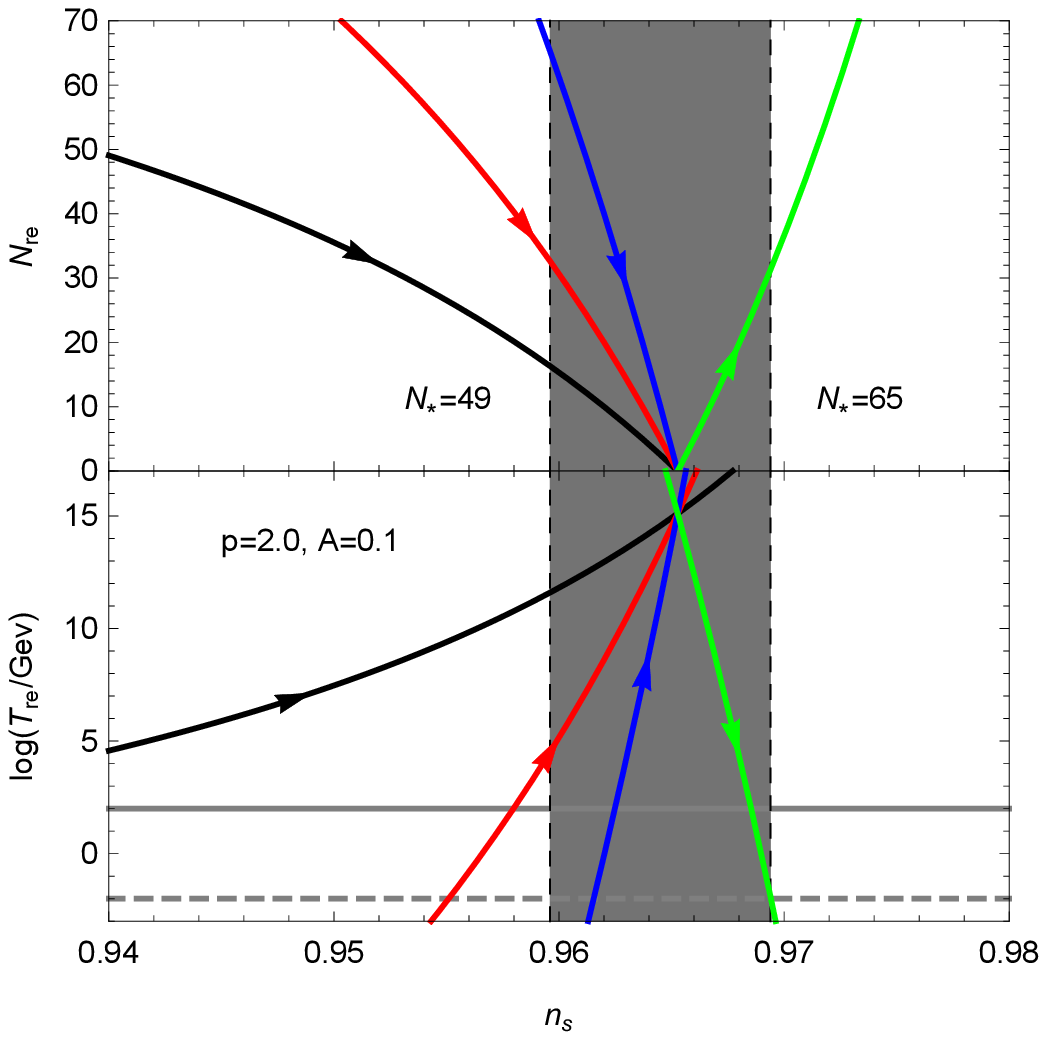}&
\includegraphics[width=0.4\textwidth]{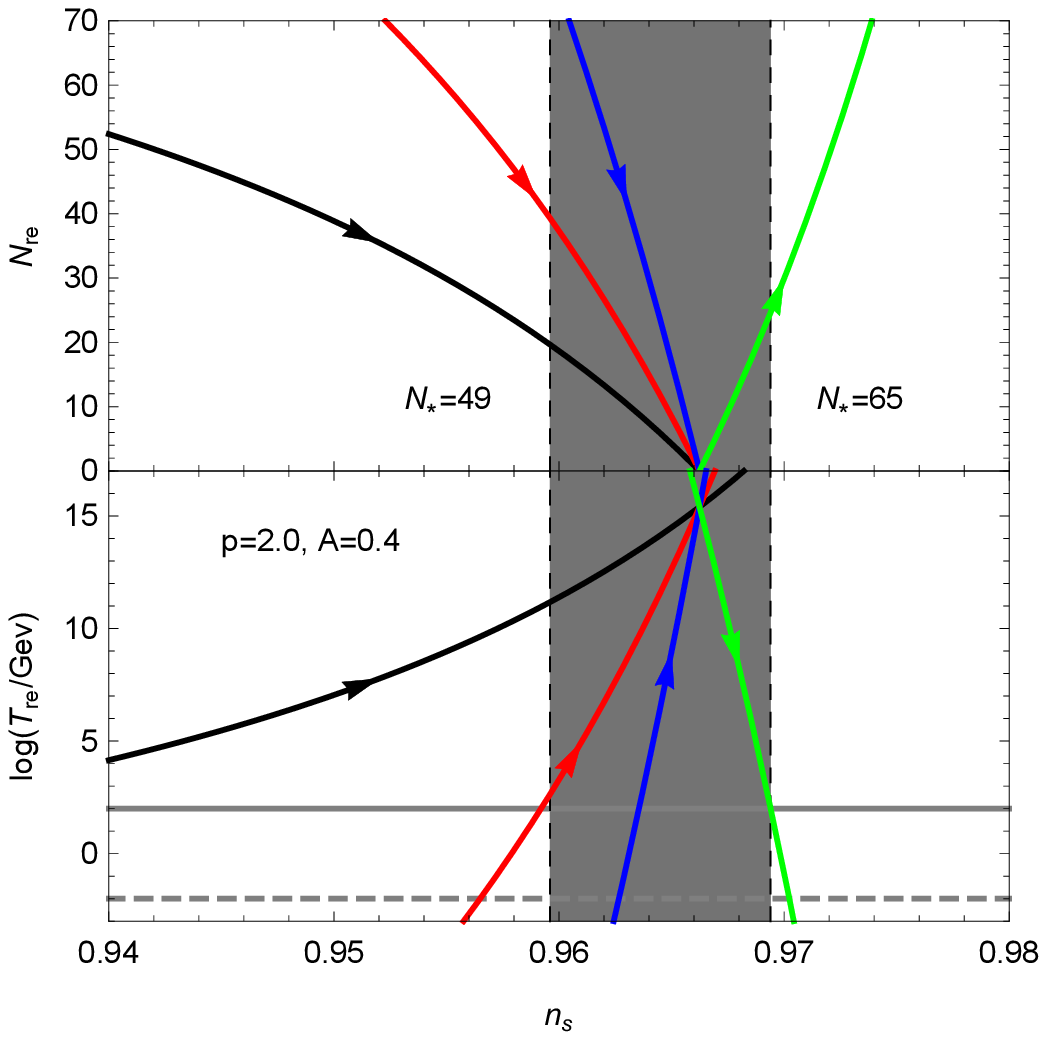}\\
\includegraphics[width=0.4\textwidth]{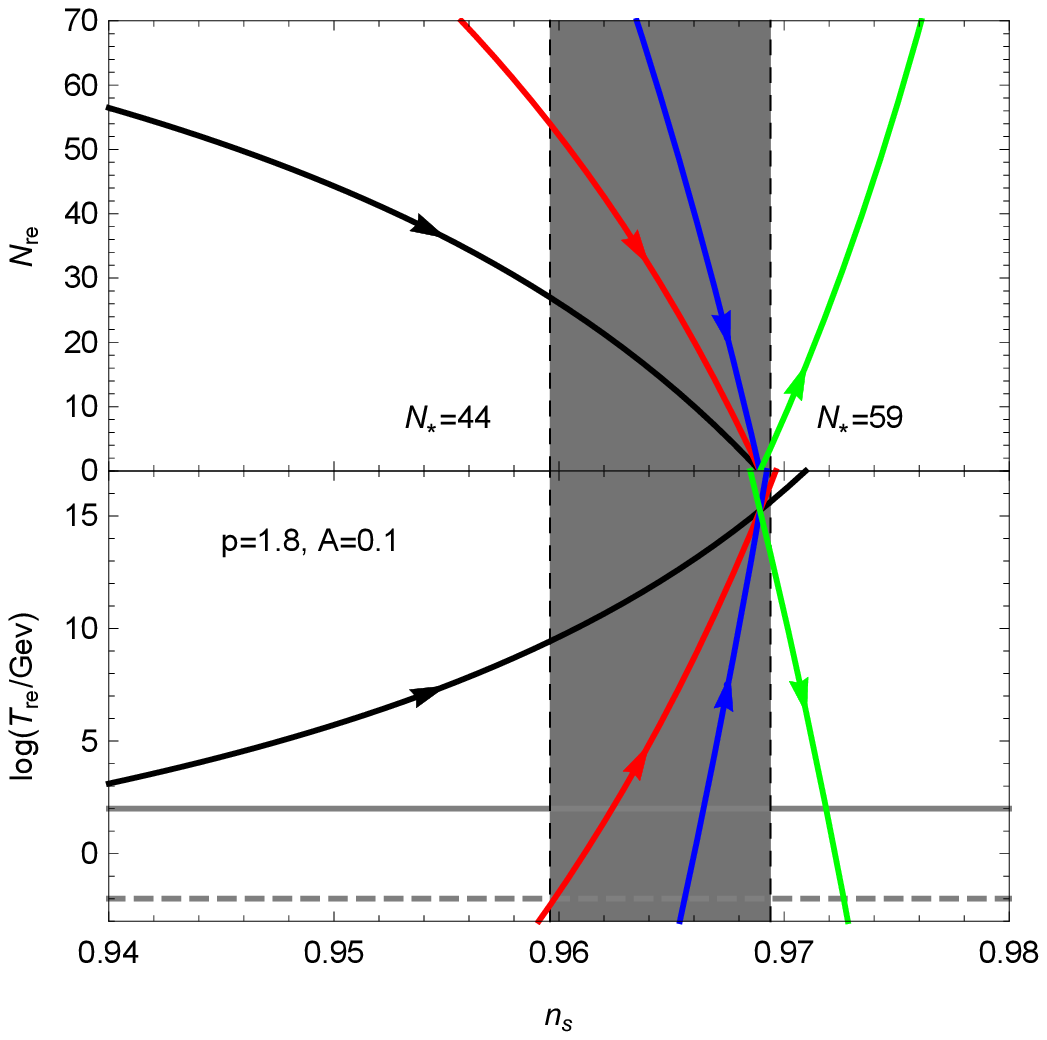}&
\includegraphics[width=0.4\textwidth]{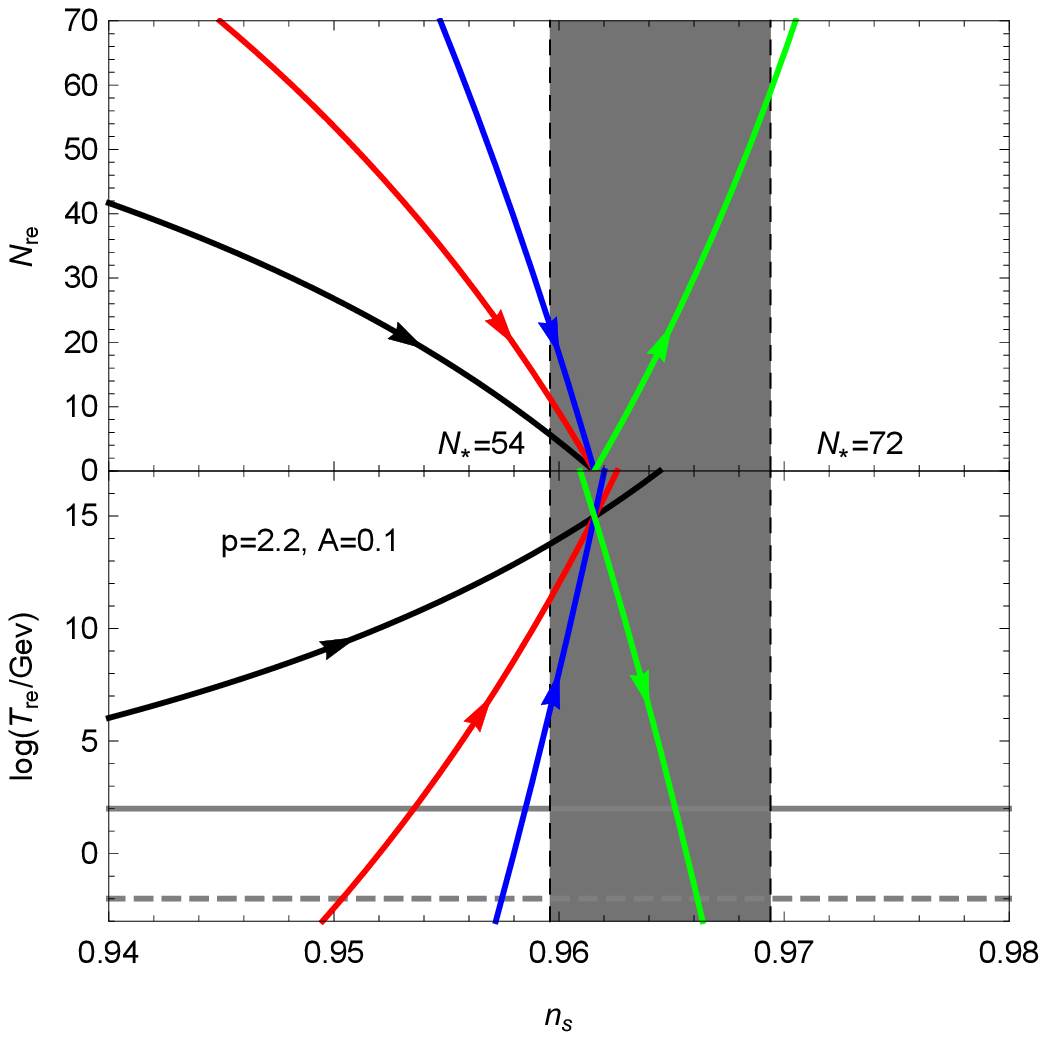}\\
\end{array}$
\caption{Same as figure \ref{fig:reheating:eta} but for the model \eqref{nspareq5}, the values of the model parameters $p$ an $A$ are indicated in each panel.}
\label{fig:reheating:PA}
\end{figure}

For the model \eqref{nspareq5},  we consider the case $p>1$ and $p\neq 1+2A$. Substituting eqs. \eqref{nspareq1} and \eqref{nspareq2} into eqs. \eqref{Nre:final} and \eqref{reheq5},  we get
\begin{gather}
\label{reheq10}
\begin{split}
  N_{re}=&\frac{4}{1-3w_{re}}\left\{60.38+\frac{1}{4}\ln\left[C(p-1)^2+(p-1)A^{1-p}\right]
  -N_\ast\right.\\
&\quad \left.-\frac{p}{4}\ln\left(N_\ast+A\right)-\frac{1}{2}\ln\left[(N_\ast+A)^{1-p}+C(p-1)\right]\right\},
\end{split}\\
\label{reheq11}
 T_{re}=0.01\left(N_\ast+A \right)^{-p/4}\left[\frac{p-1}{A^{1-p}+(p-1)C}\right]^{1/4}\exp\left[-\frac{3N_{re}(1+w_{re})}{4}\right].
\end{gather}
By choosing different values of $p$, $A$, $N_*$ and $w_{re}$, we calculate $n_s$, $N_{re}$ and $T_{re}$
from eqs. \eqref{conseq4}, \eqref{reheq10} and \eqref{reheq11}, and the results are shown in figure \ref{fig:reheating:PA}.
From figure \ref{fig:reheating:PA}, we see that depending on the model parameters $p$ and $A$ and the
value of $w_{re}$, the constraints on $N_{re}$ and $T_{re}$ are different,
but the parameter $A$ has little impact on the reheating phase.
For the parameters $p$ and $A$ that make $n_s$ consistent with the observation,
reheating with $-1/3\le w_{re}\le 2/3$ are all consistent with the observations.
As $n_s$ becomes larger, the allowed reheating epoch becomes longer for $w_{re}=-1/3$, 0 and $1/6$
while the allowed reheating epoch becomes shorter for $w_{re}=2/3$.

\begin{figure}
$\begin{array}{cc}
\includegraphics[width=0.4\textwidth]{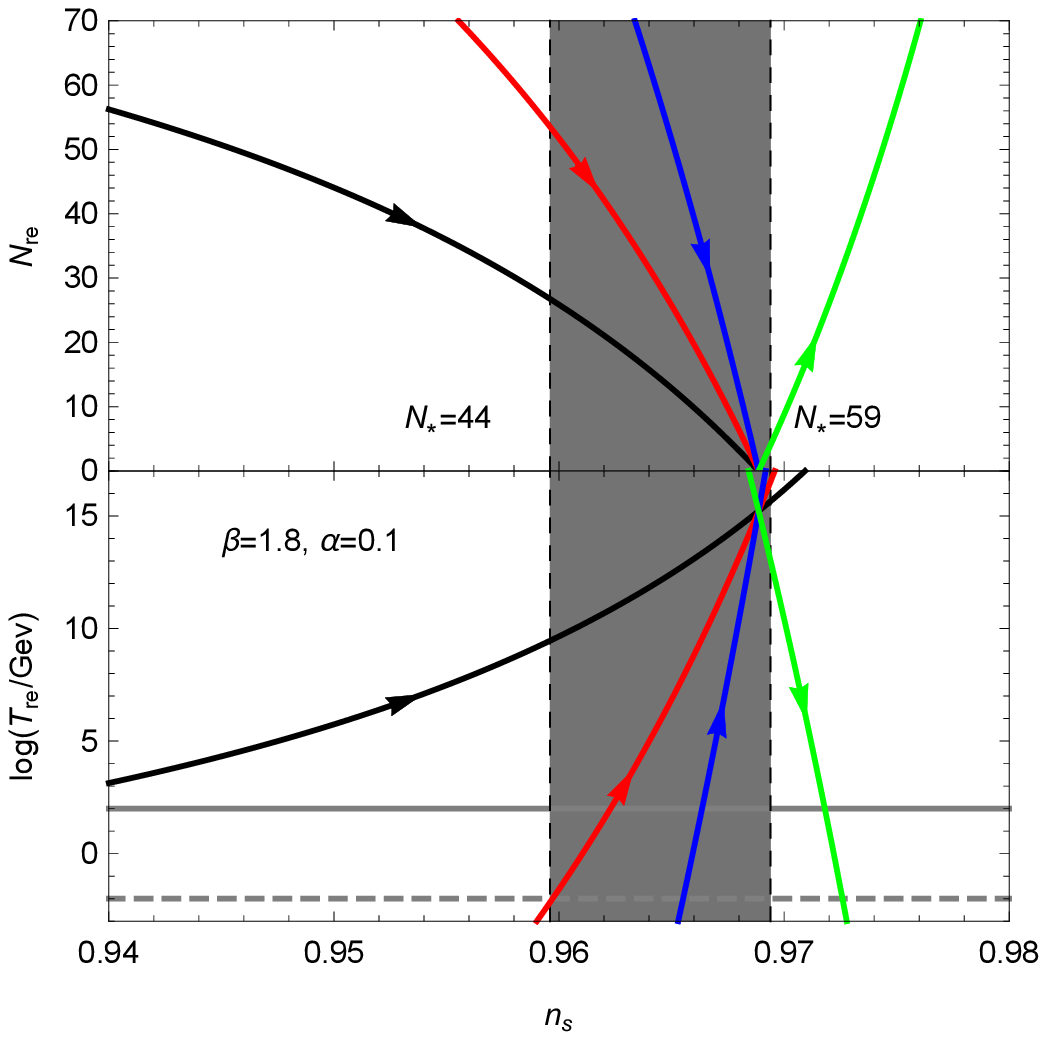}&
\includegraphics[width=0.4\textwidth]{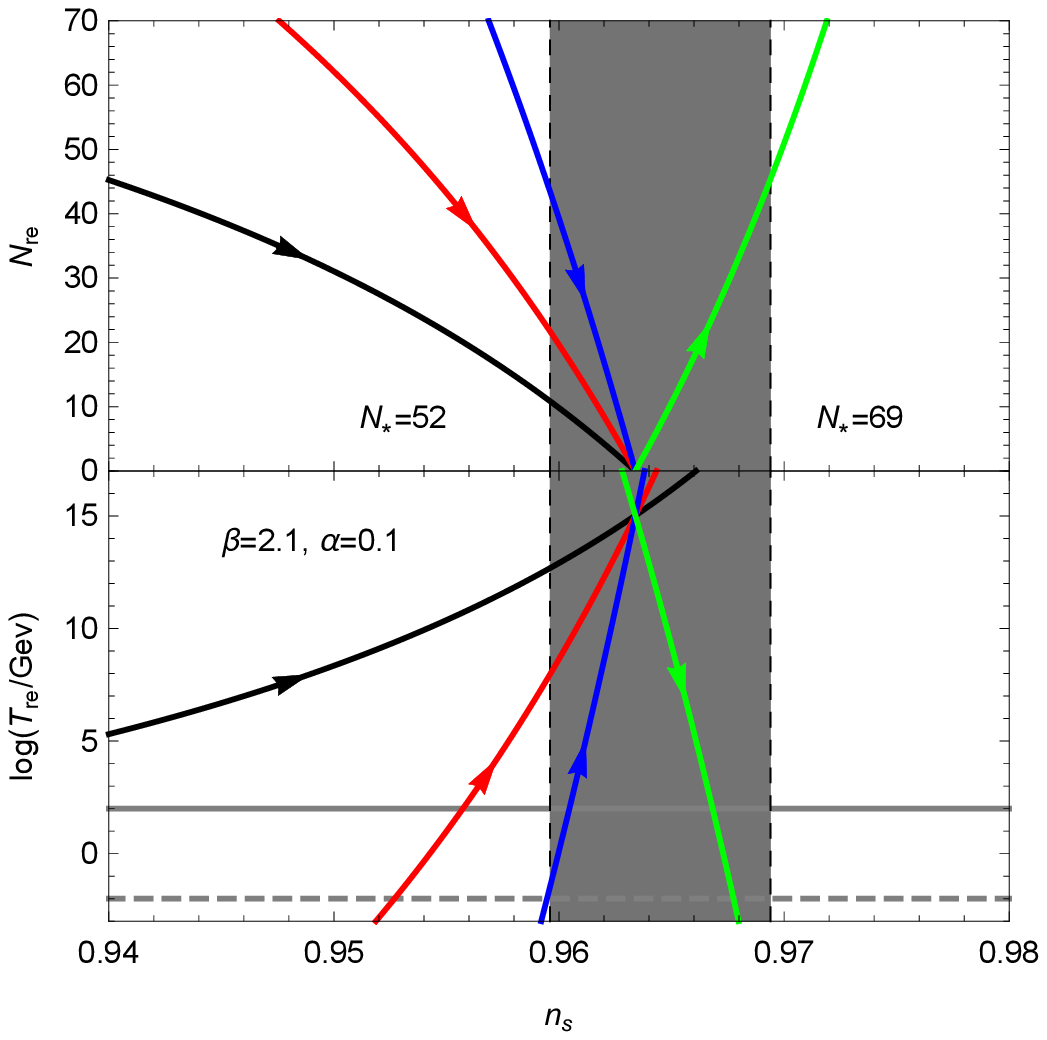}\\
\includegraphics[width=0.4\textwidth]{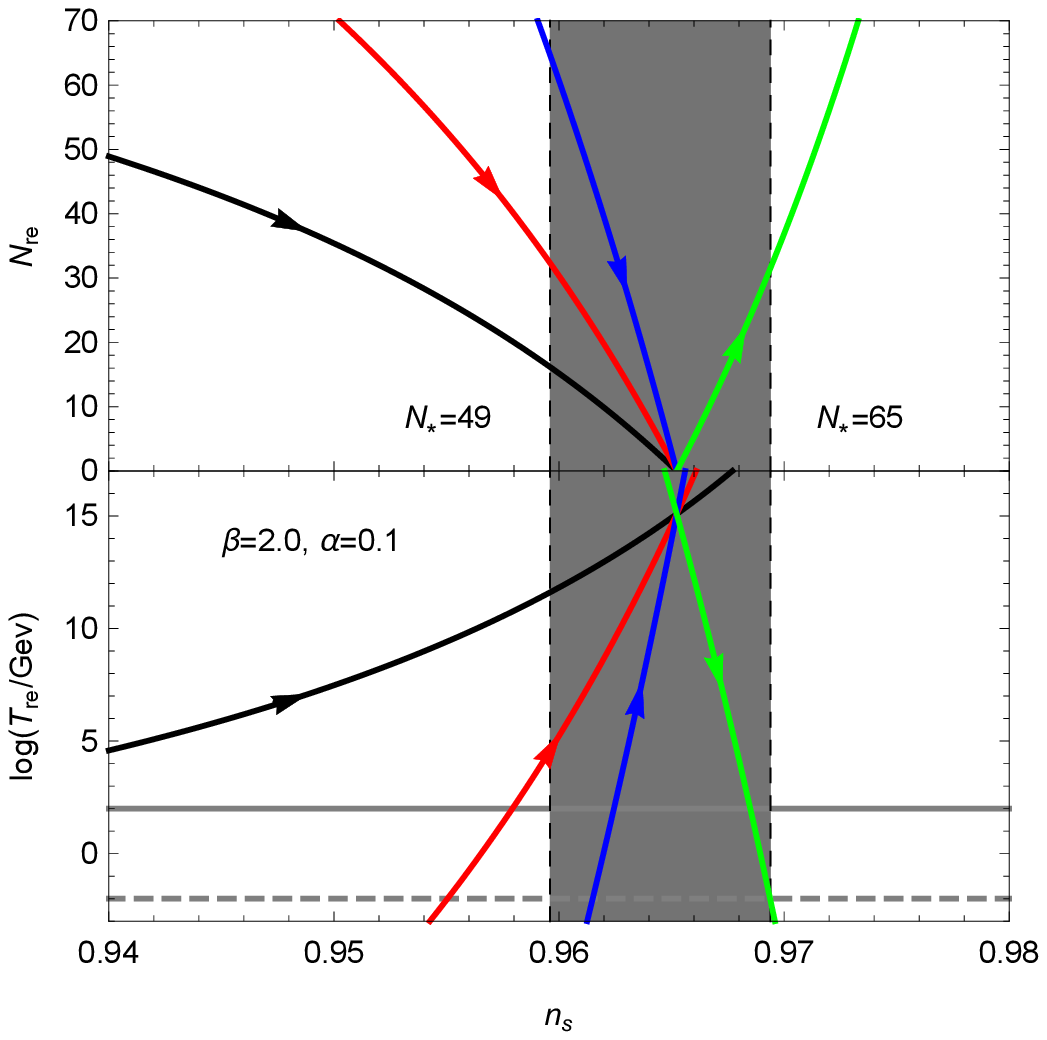}&
\includegraphics[width=0.4\textwidth]{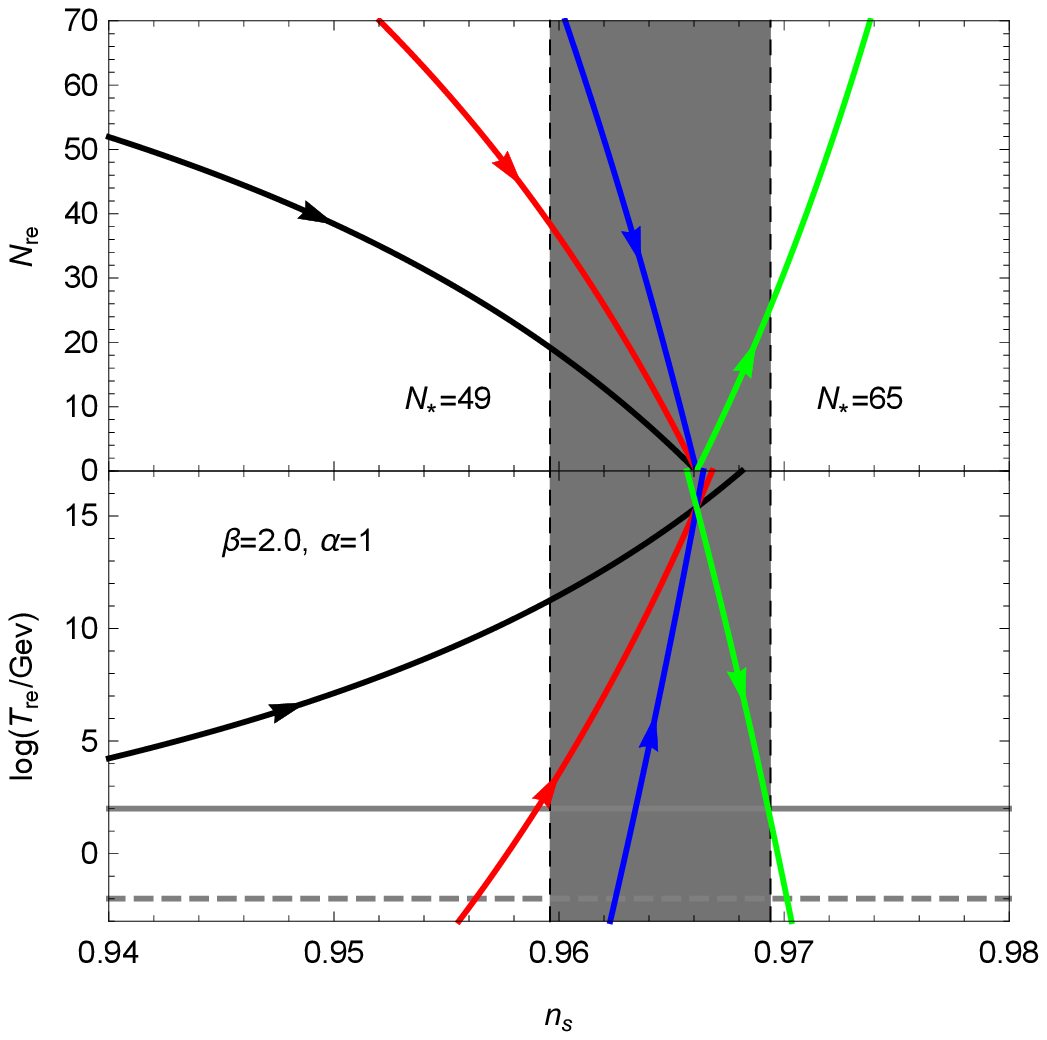}\\
\end{array}$
\caption{Same as figure \ref{fig:reheating:eta} but for the model \eqref{rpareq1}, the values of the model parameters $\alpha$ and $\beta$ are indicated in each panel.}
\label{fig:reheating:ba}
\end{figure}

For the model \eqref{rpareq1}, we consider the case $\beta\neq1$, Substituting eqs. \eqref{rpareq1} and \eqref{rpareq4} into eqs. \eqref{Nre:final} and \eqref{reheq5}, we obtain
\begin{gather}
\label{reheq12}
  N_{re}=\frac{4}{1-3w_{re}}\left[60.56+\frac{\alpha}{2(\beta-1)}-N_\ast
  -\frac{\alpha}{2(\beta-1)}\left(\frac{N_\ast+\alpha}{\alpha}\right)^{1-\beta}-\frac{\beta}{4}\ln\left(\frac{N_\ast+\alpha}{\alpha}\right)\right],\\
\label{reheq13}
  T_{re}=0.01\left(\frac{\alpha}{N_\ast+\alpha}\right)^{\beta/4}\exp\left[-\frac{3N_{re}(1+w_{re})}{4}
  +\frac{\alpha}{2(1-\beta)}\left(1-\left(\frac{N_\ast+\alpha}{\alpha}\right)^{1-\beta}\right)\right].
\end{gather}
By choosing different values of $\alpha$, $\beta$, $N_*$ and $w_{re}$, we calculate $n_s$, $N_{re}$ and $T_{re}$
from eqs. \eqref{conseq4}, \eqref{reheq12} and \eqref{reheq13}, and the results are shown in figure \ref{fig:reheating:ba}.
It is obvious that depending on the model parameters $\alpha$ and $\beta$ and the
value of $w_{re}$, the constraints on $N_{re}$ and $T_{re}$ are different,
but the parameter $\alpha$ has little impact on the reheating phase.
For the parameters $\alpha$ and $\beta$ that make $n_s$ consistent with the observation,
reheating with $-1/3\le w_{re}\le 2/3$ are all consistent with the observations.
As $n_s$ becomes larger, the allowed reheating epoch becomes longer for $w_{re}=-1/3$, 0 and $1/6$
while the allowed reheating epoch becomes shorter for $w_{re}=2/3$.

\section{Conclusions and Discussions}
Similar to the usual inflation with canonical scalar field, there is also a lower bound on the field excursion for the tachyon inflation,
but the lower bound for the tachyon field depends on $A_s$ and $N_*$. Using the observational value $\ln(10^{10} A_s)=3.094$ \cite{Ade:2015lrj},
we derive the lower bound $\Delta T\ge 1.18\times 10^5$ normalized with the reduced Planck mass $M_{pl}$ for $N_*=60$,
and the bound are supported by the three models discussed in this work.
Since the $\beta$-function $\beta(T)=-\sqrt{2\epsilon}$, so the reconstruction of the tachyon potentials from $\beta(T)$ is equivalent
to the reconstruction from the slow-roll parameter $\epsilon(T)$ or other parameterizations with the number of $e$-folds $N$.
We focus on the reconstruction of tachyon potentials from the parameterizations with $N$.

Following the reconstruction procedure presented in subsection \ref{reconsrel}, we reconstruct three classes of tachyon potentials
by parameterizing the slow-roll parameters $\epsilon$ (equivalent to the tensor to scalar ratio $r$), $\eta$ and the observable $n_s$, respectively.
We first consider the case that the slow-roll parameter is a constant, we find that only the model with $\eta_V$ being a constant is consistent with
the observations at the $1\sigma$ level, this model is therefore called the constant slow-roll inflation.
For $N_*=60$, we get $-0.0374< \eta_V< -0.0142$ at the $1\sigma$ level, $-0.0435 < \eta_V< -0.0031$ at the  $2\sigma$ level and $-0.0473 <\eta_V< 0.0067$
at the $3\sigma$ level, so the concave potential is favored by the observations at more than $2\sigma$ level.
For the simple model with $n_s=1-p/(N+A)$, the potential is either power-law or exponential form. Since the observations constrain $A<1$, so
the effect of $A$ is negligible except setting the boundary $p=1+2A$ for the parameter $p$.
The $\alpha$ attractor is the special case with $p=2$ and it is consistent with observations.
For the power-law parametrization $r=16\gamma/(N+\alpha)^\beta$, we find $\beta\sim 2$ is favored
by the observations. For all three models, if we take $n_s=0.968$ and $r=0.22$, the reconstructed potentials behave similarly and they are concave potentials.

Depending on the model parameters and the
value of $w_{re}$, the constraints on $N_{re}$ and $T_{re}$ are different,
although the parameter $A$ in the model \eqref{nspareq5} and the parameter $\alpha$ in the model \eqref{rpareq1}
have little impact on the reheating phase.
For all three models, if we choose the model parameters so that $n_s$ is consistent with the observations,
then reheating with $-1/3\le w_{re}\le 2/3$ are all consistent with the observations.
Furthermore, as $n_s$ increases, the allowed reheating epoch becomes longer for $w_{re}=-1/3$, 0 and $1/6$
while the allowed reheating epoch becomes shorter for $w_{re}=2/3$.

In summary, the main results are: (1) We derive the lower bound on the field excursion for the tachyon inflation,
which is determined by the amplitude of the scalar perturbation $A_s$ and $N_*$.
The bound is supported by all three models discussed. (2) For the models with constant slow-roll parameter,
only the model with $\eta_V$ being a constant is consistent with
the observations at the $1\sigma$ level and concave potentials are favored by the observations.
(3) As $n_s$ increases, the allowed reheating epoch becomes longer for $w_{re}=-1/3$, 0 and $1/6$
while the allowed reheating epoch becomes shorter for $w_{re}=2/3$.

\acknowledgments
This research was supported in part by the National Natural Science
Foundation of China under Grant No. 11475065 and
the Major Program of the National Natural Science Foundation of China under Grant No. 11690021.


\providecommand{\href}[2]{#2}\begingroup\raggedright\endgroup

\end{document}